\documentclass[12pt]{article}

\begin{document}

\newcommand{\df}{\stackrel{\rm def}{=}}
\newcommand{\co}{{\scriptstyle \circ}}
\newcommand{\lb}{\lbrack}
\newcommand{\rb}{\rbrack}
\newcommand{\rn}[1]{\romannumeral #1}
\newcommand{\msc}[1]{\mbox{\scriptsize #1}}
\newcommand{\dsp}{\displaystyle}
\newcommand{\scs}[1]{{\scriptstyle #1}}

\newcommand{\ket}[1]{| #1 \rangle}
\newcommand{\bra}[1]{| #1 \langle}
\newcommand{\vac}{| \mbox{vac} \rangle }

\newcommand{\e}{\mbox{{\bf e}}}
\newcommand{\va}{\mbox{{\bf a}}}
\newcommand{\bc}{\mbox{{\bf C}}}

\newcommand{\com}{C\!\!\!\!|}

\newcommand{\br}{\mbox{{\bf R}}}
\newcommand{\bz}{\mbox{{\bf Z}}}
\newcommand{\bq}{\mbox{{\bf Q}}}
\newcommand{\bn}{\mbox{{\bf N}}}
\newcommand {\eqn}[1]{(\ref{#1})}

\newcommand{\cp}{\mbox{{\bf P}}^1}
\newcommand{\n}{\mbox{{\bf n}}}
\newcommand{\sbz}{\msc{{\bf Z}}}
\newcommand{\sn}{\msc{{\bf n}}}

\newcommand{\be}{\begin{equation}}\newcommand{\ee}{\end{equation}}
\newcommand{\bea}{\begin{eqnarray}} \newcommand{\eea}{\end{eqnarray}}
\newcommand{\ba}[1]{\begin{array}{#1}} \newcommand{\ea}{\end{array}}

\newcommand{\cleqn}{\setcounter{equation}{0}}

\makeatletter

\@addtoreset{equation}{section}

\def\theequation{\thesection.\arabic{equation}}
\makeatother

\def\np{Nucl. Phys. {\bf B}}\def\pl{Phys. Lett. {\bf B}}
\def\mpl{Mod. Phys. {\bf A}}\def\ijmp{Int. J. Mod. Phys. {\bf A}}
\def\cmp{Comm. Math. Phys.}\def\prd{Phys. Rev. {\bf D}}

\def\oa{\bigcirc\!\!\!\! a}
\def\ob{\bigcirc\!\!\!\! b}

\def\ve{\vec e}\def\vk{\vec k}\def\vn{\vec n}\def\vp{\vec p}
\def\vr{\vec r}\def\vs{\vec s}\def\vt{\vec t}\def\vu{\vec u}
\def\vv{\vec v}\def\vx{\vec x}\def\vy{\vec y}\def\vz{\vec z}


\begin{flushright}
La Plata Th/03-02\\May, 2004
\end{flushright}

\bigskip

\begin{center}

{\Large\bf About the stability of a $D4$-$\bar D 4$ system}
\footnote{
This work was partially supported by CONICET, Argentina
}
\bigskip
\bigskip

{\it \large Adri\'{a}n R. Lugo}
\footnote{
{\sf lugo@fisica.unlp.edu.ar}
}
\bigskip

{\it Departamento de F\'\i sica, Facultad de Ciencias Exactas \\ Universidad Nacional
de La Plata\\ C.C. 67, (1900) La Plata, Argentina}
\bigskip
\bigskip

\end{center}
\bigskip

\begin{abstract}
We study a system of coincident $D4$ and $\bar D 4$ branes with non zero world-volume
magnetic fields in the weak coupling limit.
We show that the conditions for absence of tachyons in the spectrum coincide exactly with
those found in hep-th/0206041, in the low energy effective theory approach, for the system
to preserve a $\frac 14$ of the supersymmetries of the Type IIA string theory vacuum.
We present further evidence about the stability of the system by computing the lowest order
interaction amplitude from both open and closed channels, thus verifying the no force
condition as well as the supersymmetric character of the spectrum.
A brief discussion of the low energy effective five dimensional world-volume theory is given.
\end{abstract}

\bigskip
\section{Introduction}
\cleqn

The discovery, in type II string theories, of cylinder-like branes preserving a
quarter of the supersymmetries of the flat Minkowski space-time, the so-called
``supertubes", has attracted much attention recently \cite{sutubo1}-\cite{nobu1}.
The stabilizing factor at the origin of their BPS character, which prevents them from
collapsing, is the angular momentum generated by the non-zero gauge field that lives on
the brane.

An interesting feature of the supertube is that it presents $D0$ and $F1$ charges, but no
$D2$ charge.
In relation to this fact, Bak and Karch (BK) observed that, if we consider the elliptical
supertube in the limit when one of the semi-axis  goes to infinity, the resulting system
should be equivalent to having two flat $2$-branes with total $D2$ charge equal to zero.
This naturally led to conjecture the existence of SUSY $D2$-$\bar D2$ systems.
Such a study as well as the study of systems with arbitrary numbers of $D2$ and $\bar D2$
branes was made in the context of the Born-Infeld action in reference \cite{bakar}, where
the conditions to be satisfied by the Killing spinors were identified.
Soon after that, in reference \cite{lugo1}, higher dimensional brane-antibrane systems were
considered  in the Born-Infeld context (see \cite{nobu2} for related work in the matrix
model context).
In particular, the existence of a quarter SUSY $D4$-$\bar D4$ systems
with $D2$ and Taub-NUT charges and no $D4$-brane charge that should represent genuine
bound states of such components was conjectured.
While it is plausible that a five dimensional supertube-like solution exists, leading in a
certain limit to the brane-antibrane system, much as it happens with the supertube,
in this paper we will focus on a detailed study of the conformal field theory and,
in particular, in the absence of tachyonic instabilities in the system (in the supertube
context such analysis was carried out in references \cite{sutubo2}, \cite{bakota}).
\bigskip

\section{Review of the construction of the $D4$-$\bar D 4$ SUSY system}
\cleqn

Let us start by remembering those results in \cite{lugo1} which are relevant to the
subject studied in this paper.

Let us consider type $IIA$ superstring theory in the flat vacuum defined by the ten
dimensional Minkowskian metric tensor $G = \eta_{MN}dX^M\,dX^N$, with constant dilaton
and the other fields put to zero.
This background preserves maximal SUSY, whose general Killing spinor is, in the standard local
basis $dX^M$, a constant $32$ dimensional Majorana spinor $\epsilon$.
Let us consider a $D4$ (or $\bar D4$) brane, with world-volume coordinates
$\{\xi^\mu,\mu=0,1,2,3,4\}$, the embedding defined by $X^\mu(\xi) = \xi^\mu, \,
\mu=0,\dots,4\;,\;X^i(\xi)=0\; ,\; i=5,\dots,9$, and an abelian gauge field $A_\mu(\xi)
= \frac{1}{2}\, F_{\nu\mu}\,\xi^\nu\,$ living on the brane, being $F=dA$ the constant field
strength.
Then, the introduction of such a $D4 $ brane in space-time will preserve
the supersymmetries that satisfy \cite{berg2}
\footnote{
The scale $T_s= (2\,\pi\,\alpha')^{-1}$ is put to unity everywhere unless explicitly written.
}
\be
\Gamma\;\epsilon = \pm\;\epsilon\;\;\;\;, \label{susycond}
\ee
where the ``-" sign on the r.h.s.
corresponds to the $\bar D 4$ brane with the {\it same} fields as the $D4$ brane
because, by definition, it has opposite orientation to the $D4$ brane.
This last orientation is defined by $\varepsilon_{01234} =+1$, which is present in the
$\Gamma$-matrix \cite{berg1}
\footnote{
We take the ten dimensional $\Gamma$-matrices to obey
$\{\Gamma^M\,;\Gamma^N\} = 2\,\eta^{MN}\;$, $\,\{\Gamma^M\,;\Gamma_{11}\} =0 \;$, with
$\Gamma_{11}\equiv\Gamma^1\dots\Gamma^9\;\Gamma^0$.
For definiteness, we adopt a Majorana-Weyl basis where
$\Gamma^M{}^t = \eta_{MM}\,\Gamma^M\;$.
In such basis, we can take $A_\pm = C_\pm\;$,
where $\;C_\pm\; \Gamma^M \; C_\pm{}^{-1} = \pm\Gamma^M{}^\dagger =\pm\Gamma^M{}^t\;$
defines the charge conjugation matrices.
}
\bea
\Gamma &\equiv& d^{-\frac{1}{2}}\; \left( \Gamma_{11} + \frac{1}{2}
F_{\mu\nu}\;\Gamma^{\mu\nu}\; + \frac{1}{8}F_{\mu\nu}\;F_{\rho\sigma}\;
\Gamma^{\mu\nu\rho\sigma}\;\Gamma_{11}\right)\;\Gamma^{01234}\cr d&\equiv& \det d_+ =
\det d_-> 0
\;\;\;\;,\;\;\; d_\pm{}^\mu{}_\nu = \delta^\mu{}_\nu + F^\mu{}_\nu \label{gama1}
\eea

Now, as discussed in \cite{lugo1}, the anti-symmetric matrix of  magnetic fields
$(B_{ij}= F_{ij})$, with $\, i,j=1,\dots,4\,,$ can be put in the standard form
$\left(\matrix{B_1&0\cr 0&B_2}\right)\otimes i\sigma_2$ by means of an $SO(4)$-rotation;
the $SO(2)\times SO(2)$ rotation left over by this condition can then be used to put the
electric field $E_i\equiv F_{i0}$ in the $(13)$ plane.
So we can consider, with no loss of generality
\be
(F^\mu{}_\nu) = \left(\matrix{ 0  & E_1 &  0  & E_3 & 0\cr
                              E_1 & 0   & B_1 &  0   & 0\cr
                               0  &-B_1  & 0  &  0   & 0\cr
                              E_3 & 0   &  0  &  0   & B_2\cr
                               0  & 0   &  0  & -B_2 & 0 \cr}\right)\;\;\;\;,\label{fd4}
\ee
because any other solution will be related to the ones with the configuration
(\ref{fd4}) by means of successive rotations.

One of the solutions found in \cite{lugo1} corresponds to a T-dual configuration of the
Bak-Karch $D2-\bar D 2$ system; we will be interested at present on another one.
This novel solution, which also preserves $\frac{1}{4}$ of SUSY, is obtained by restricting
the field strength in the following way
\be
B_1{}^2\,B_2{}^2 - B_1{}^2\, E_3{}^2 - B_2{}^2\,E_1{}^2 = 1 \label{const42}
\ee
This constraint clearly implies that $(B_{ij})$ cannot be singular; even more, if it
holds, then the module of the four-vector $\vec\beta = - B^{-1}\,\vec E$ verifies
\be
0<\beta^2 = \left(\frac{E_1}{B_1}\right)^2 + \left(\frac{E_3}{B_2}\right)^2
=1-\frac{1}{B_1{}^2\,B_2{}^2} <1\;\;\;\; .
\ee
It is possible to show that a boost with a velocity equal to $\vec\beta$ eliminates the
electric field; a further rotation (which certainly does not affect the null electric field
condition) can fix the magnetic fields matrix in the standard form again.
So we conclude that the sector of fields obeying (\ref{const42}) is Lorentz-equivalent to a
sector of the observers that do not see electric field and have non-singular magnetic matrix
of determinant equal to one.
Therefore, we will restrict our attention to a field strength (\ref{fd4}) with
\bea
E_1&=&E_3=0\cr
\det B &=& B_1{}^2\,B_2{}^2=1\;\;\;\; .\label{susycons}
\eea
Such field strength clearly breaks the space-time symmetry as follows
\be
SO(1,9) \longrightarrow \br\times SO(2)\times SO(2)\times SO(5)\;\;\;\; .
\ee

The $\Gamma$-matrix (\ref{gama1}) takes the form
\bea
\Gamma &=&d^{-\frac{1}{2}} \;\left(
\Gamma_{01234}\;\Gamma_{11} + B_1\,\Gamma_{034} + B_2\,\Gamma_{012} +
B_1\;B_2\;\Gamma_0\;\Gamma_{11}\right)\cr d &=& 2 + B_1{}^2 + B_2{}^2 = ( |B_1 | +
|B_2|)^2 \;\;\;\; .\label{gama4}
\eea
The solution we are interested in is obtained by splitting equation (\ref{susycond})
into the following two conditions
\vfill\eject
\bea
- B_1\,B_2\;\Gamma_{1234}\;\epsilon &=& \epsilon\cr d^{-\frac{1}{2}}\; \left( B_1\;
\Gamma_{034} + B_2\;\Gamma_{012}\right)\;\epsilon &=& \pm\epsilon\;\;\;\; .\label{condb42}
\eea
This system is consistent, and leads to the preservation of $\frac{1}{4}\,32 = 8$ SUSY
charges; explicitly, with $-i\Gamma_{12}(s)= (-)^s\,(s)\;$, etc (see \cite{lugo1} for
details and notation), the $8$ Killing spinors are
\be
\eta^{(\pm)}_{(s_1s_2s_3)} = (s s_1s_2s_3 1) \pm i\, sg(B_2)\; (-)^{\sum_{k=1}^3 s_k}\;
(s s_1 s_2s_3 0)\;\;\;,\;\;\; (-)^s = sg(B_1\,B_2)\;(-)^{s_1}\;\; , \label{kill42}
\ee
where the labels $s$ take the values $0$ (spin down) or $1$ (spin up).
From these equations we can see that the Killing
spinors $\eta^{(-)}_{(s_1s_2s_3)}$ corresponding to the introduction of a $\bar D 4$
brane with world-volume gauge fields $(-B_1,-B_2)$ are exactly the same as the Killing
spinors $\eta^{(+)}_{(s_1s_2s_3)}$ of the $D4$-brane with fields $(B_1, B_2)$.
Therefore, we are led to conjecture that the $\eta^{(+)}_{(s_1s_2s_3)}$ in
(\ref{kill42}) are Killing spinors (without taking into account back-reaction effects)
of $\frac{1}{4}$ SUSY systems of $D4$ branes with fields $(B_1,B_2)$ and $\bar D4$
branes with the opposite ones $(-B_1, -B_2)$, obeying $B_1{}^2\,B_2{}^2 =1$.
In what follows, we will focus our attention on a $D4$-$\bar D4$ configuration.

\bigskip
\section{The ``light-cone" gauge-fixing and rotational invariance}
\cleqn

As is well known, a covariant analysis of the string spectrum is carried out by
considering the BRST charge \cite{polcho2}
\bea
Q^{BRST} &\equiv& \sum_{m\in\bz}\; c_{-m}\; L_m^{(m)} +\sum_{r\in\bz_\delta}\;
\gamma_{-r}\; G_r^{(m)}
+ \frac{1}{2}\; \sum_{m\in\bz}\; :c_{-m}\; L_m^{bc} : + \frac{\Delta_0^{bc}}{2}\; c_0\cr
&+& \sum_{m\in\bz}\; c_{-m}\; L_m^{\beta\gamma} - \sum_{m\in\bz}\,\sum_{r\in\bz_\delta}\;
b_m\;\gamma_{r-m}\;\gamma_{-r}\;\;\;\; ,
\eea
where ($\beta$-$\gamma$) $b$-$c$ are the ($\lambda = \frac{3}{2}$ super) $\lambda=2$ ghost
fields and the superscript ``(m)" stands for ``matter".
It verifies
\be
\{Q^{BRST};\,Q^{BRST}\} = \frac{c^{(m)} + c^{(g)}}{12}\; \left(
\sum_{m\in\bz}\; m\,(m^2-1)\; c_{-m}\; c_m  + \sum_{r\in\bz_\delta}\;(4r^2 -1)\;\gamma_{-r}\;
\gamma_r\right)\;\; .\label{qn}
\ee
From (\ref{qn}), it follows that $Q^{BRST}$ is nilpotent iff the central charge of the matter
system (whatever it is) is $\; c^{(m)} = -c^{(g)}= 15\;$, where we have used (\ref{ccsg}).
Physical states are then defined as cohomology classes of this operator.
However for the sake of clarity we will analyze the spectrum in the light-cone gauge
to be described in what follows.

Let us start with a brief review of the analogue of the light-cone gauge fixing
procedure in the presence of branes, a subject which, to our understanding, is not
covered deeply enough in the literature.
Let us consider an open superstring theory that consists of a time-like coordinate $X^0$
with NN b.c., $d+1$ coordinates $X^I, I=1,\dots,d+1$ with homogeneous DD b.c.
(fermionic partners are considered below) and an arbitrary $N=1$ superconformal field theory.
Let us pick up $X^0$ and $X^{d+1}$ to define ``light-cone" coordinates
$X^\pm\equiv X^0\pm X^{d+1}$.
Then, the b.c. are equivalently written as
\be
\left.\begin{array}{r}
\partial_\sigma X^0 |_{\sigma = 0} = \partial_\sigma X^0 |_{\sigma = \pi}= 0\cr
\partial_\tau X^{d+1} |_{\sigma = 0}=\partial_\tau X^{d+1} |_{\sigma = \pi}= 0
\end{array}\right\}\longleftrightarrow
\left\{\begin{array}{l}
\partial_\pm X^+ |_{\sigma = 0}=\partial_\mp X^- |_{\sigma = 0}\cr
\partial_\pm X^+ |_{\sigma = \pi}=\partial_\mp X^- |_{\sigma = \pi}
\end{array}\right.\;\;\; ,\label{lcbc}
\ee
which show that, although it could seem at first sight a little bizarre to mix
fields with different b.c., there is no obstacle to doing so, and the problem on the r.h.s
of (\ref{lcbc}) is perfectly well-defined; the same happens if $X^{d+1}$ is ND or DN.
What changes radically is the interpretation of the complete gauge fixing possible
thanks to the (super) conformal invariance of the string theory.
From (\ref{nn}), (\ref{dd}) we get for the light-cone coordinates
\bea
X^\pm(\tau,\sigma) &=& x^\pm + X_L^\pm (\sigma^+) +
X_R^\pm (\sigma^-)\cr X_L^\pm(\sigma^+) &=& \alpha'\;p^\pm\;\sigma^+ +
i\,\frac{l}{2}\;\sum_{m\in\bz'}\; \frac{\alpha_m^\pm}{m} \; e^{-im\sigma^+}\cr
X_R^\pm(\sigma^-) &=& \alpha'\;p^\mp\;\sigma^- + i\,\frac{l}{2}\;\sum_{m\in\bz'}\;
\frac{\alpha_m^\mp}{m} \; e^{-im\sigma^-}\;\;\;\; ,\label{lcexp1}
\eea
where $\;x^\pm \equiv x^0\pm x_0^{d+1}=x^0\;$ and
$p^\pm\equiv p^0\pm T_s\,\Delta x^{d+1}=p^0$.
Now, in order to linearize the $L_0$ constraint, a reparameterization very different from
the usual one (see for example \cite{GSW1}) must be considered.
Let us try in fact, the one defined by
\bea
\tilde\tau(\tau,\sigma) &=& \frac{1}{2\,\alpha'\,p^+}\;\left( X_L^+(\sigma^+) +
X_R^-(\sigma^-) \right) = \tau + \frac{i}{l\,p^+} \sum_{m\in\bz'}\;
\frac{\alpha_m^+}{m} \; e^{-im\tau}\;\cos m\sigma\cr \tilde\sigma(\tau,\sigma) &=&
\frac{1}{2\,\alpha'\,p^+}\;\left( X_L^+(\sigma^+) - X_R^-(\sigma^-) \right) = \sigma +
\frac{1}{l\,p^+} \sum_{m\in\bz'}\; \frac{\alpha_m^+}{m} \; e^{-im\tau}\;\sin m\sigma
\;\; .\cr
& &
\eea
This world-sheet diffeomorphism satisfies the crucial properties
\begin{itemize}
\item It is a conformal reparameterization;
\item It preserves the region of the parameters
$\;\tilde\tau \in\br\;,\; 0\leq\tilde\sigma\leq\pi\,$;
\item It leaves the b.c. (\ref{lcbc}) invariant.
\end{itemize}
In terms of these new parameters we get
\vfill\eject
\bea
\tilde X^+(\tilde\tau,\tilde\sigma)
&\equiv& X^+(\tau,\sigma) = x^+ + \alpha'\, p^+\;\tilde\sigma^+ + \tilde
X_R^+(\tilde\sigma^-)\cr \tilde X^-(\tilde\tau,\tilde\sigma) &\equiv& X^-(\tau,\sigma)
= x^- + \alpha'\, p^+\;\tilde\sigma^- + \tilde X_L^-(\tilde\sigma^+)\;\;\;,\label{lcexp2}
\eea
where
\bea
\tilde X_R^+(\tilde\sigma^-) &\equiv&
X_R^+(\sigma^-)|_{\sigma^-(\tilde\sigma^-)} = \alpha'\;\tilde p^-\;\tilde\sigma^- +
i\,\frac{l}{2}\;\sum_{m\in\bz'}\; \frac{\tilde\alpha_m^-}{m} \;
e^{-im\tilde\sigma^-}\cr \tilde X_L^-(\tilde\sigma^+) &\equiv&
X_L^-(\sigma^+)|_{\sigma^+(\tilde\sigma^+)} = \alpha'\;\tilde p^-\;\tilde\sigma^+ +
i\,\frac{l}{2}\;\sum_{m\in\bz'}\; \frac{\tilde\alpha_m^-}{m} \; e^{-im\tilde\sigma^+}
\eea
define the new variables $\{\tilde\alpha_m^-\}$ in terms of the old ones
$\{\alpha_m^-\}$.
From (\ref{lcexp1}) and  (\ref{lcexp2}) we see that the
reparameterization just puts $\{\tilde\alpha_m^+=0\;,\;m\neq 0\}$ (in particular $\tilde
p^- = p^-=p^0$) and lefts over a translation invariance in $\tilde\tau$, as in the usual
case; however, $\tilde X^+$ {\it is not} the world-sheet time.

Analogously, for the fermionic superpartners $\psi^0$ and $\psi^{d+1}$ we introduce
$\psi^\pm = \psi^0 \pm\psi^{d+1}$.
But, while in the usual case we can reach the gauge $\psi^+ =0$ through a superconformal
transformation (\ref{SUSY}), it turns out that in our setting the gauge fixing allows
to put $\psi_+^+= \psi_-^- = 0 \leftrightarrow b_r^+=0$
through a superconformal transformation defined by the parameters
\be
\epsilon^\pm (\sigma^\pm) = \mp\frac{\psi_\pm^\pm}{2\,\partial_\pm X^\pm}\;\;\;\; .
\ee
It is easy to check that this gauge fixing is compatible with (\ref{lcexp2}).
The super-Virasoro constraints are readily solved in this gauge; from (\ref{vir}),
(\ref{scurr})
\bea
T_{\pm\pm}(\sigma^\pm) &=& T^\perp_{\pm\pm}(\sigma^\pm) -
\partial_{\pm}X^+\;\partial_{\pm}X^- =  T^\perp_{\pm\pm}(\sigma^\pm) -
\alpha'\,\frac{l\,p^+}{2}\;\sum_{m\in\bz}\;\alpha_m^-\; e^{-im\sigma^\pm}\cr
G_\pm(\sigma^\pm) &=& G_\pm^\perp(\sigma^\pm) - \frac{1}{2}\;\psi_\pm^\mp\;\partial_\pm
X^\pm = G_\pm^\perp(\sigma^\pm) - \frac{\alpha'\,l\,p^+}{2\,\sqrt{2}}\;
\sum_{r\in\bz_\epsilon}\,b_r^-\;e^{-ir\sigma^\pm}\;\; .
\eea
It follows that the conditions $\;L_m - A\,\delta_{m,0}= G_r =0\;$ yield
\be
\alpha_m^- = \frac{2}{l\,p^+}\;(L_m^\perp - A\,\delta_{m,0})\;\;\;,\;\;\; b_r^- =
\frac{2}{l\,p^+}\; G_r^\perp\label{energy}
\ee
respectively.
Here, $A$ is a normal ordering constant, while $``{}^\perp"$ stands for contributions
other than the $(0,d+1)$ directions, i.e. the $(1,\dots, d)$ directions and the $N=1$ theory.
In particular, $(L_m^\perp , G_r^\perp)$ generate an $N=1$ superconformal algebra (\ref{n1alg})
with $\;c^\perp = \frac{3}{2}\,d + c_{N=1}\;$.

This gauge fixing obscures the initial $SO(d+1)$ invariance of the system, leaving
just the $SO(d)$ subgroup manifest.
The tentative generators $\;\{J_{IJ}\}= \{J_{ij}, J_{i(d+1)}\equiv J_i\}\;$
in the gauge-fixed system are
\vfill\eject
\bea
J_{ij} &\equiv&J^{(0)}_{ij} + J_{ij}^{(b)} + J_{ij}^{(f)}\cr &=&\frac{1}{2\,i}\,
[b_0^i\, ;b_0^j\,]-i\,\sum_{m>0}\,\frac{1}{m}\;\left( \alpha_{-m}^i\;\alpha_m^j -
 \alpha_{-m}^j\;\alpha_m^i\right) -i\,\sum_{r>0}\,\left( b_{-r}^i\; b_r^j-b_{-r}^j\;
 b_r^i\right)\cr
J_i&\equiv&J^{(0)}_{i} +  J_i^{(b)} + J_i^{(f)}\cr &=&\frac{i}{l\,p^+}\;\left(
b_0^i\;G_0^\perp + \sum_{m>0}\,\frac{1}{m}\;\left( \alpha_{-m}^i\;L^\perp_m-
L^\perp_{-m}\;\alpha^i_m\right) +\sum_{r>0}\,\left( b^i_{-r}\; G^\perp_r
-G^\perp_{-r}\; b^i_r \right)\right)\;\; ,\cr & &\label{lcgen}
\eea
where $i,j=1,\dots,d\;$ and $J^{(0)}_{ij}$ and $J^{(0)}_{i}$ are absent in the NS sector.
It is straightforward to show that
\be
[J_{ij} ;J_{kl}] = -i\,\left( \delta_{il}\; J_{jk} + \delta_{jk}\; J_{il} -
(i\leftrightarrow j) \right) \;\;\;,\;\;\; [J_{ij} ;J_k] = i\,\left( \delta_{ik}\; J_j
- \delta_{jk}\; J_i \right)
\ee
while, as usual, the problematic commutator is $[J_i ; J_j]$, whose careful computation
gives
\bea
[J_i; J_j] &=& i\,J_{ij}\;\frac{p^-}{p^+} + \frac{i}{(l\,p^+)^2}\;
(2\,A - \frac{c^\perp}{12})\; J^{(0)}_{ij}\cr &+& \frac{1}{(l p^+)^2}\;
\left(\sum_{m>0}\;\frac{\Delta_m^{(b)}}{m}\; \left( \alpha_{-m}^i\;\alpha_m^j-
(i\leftrightarrow j)\right) +\sum_{r>0}\;\Delta_r^{(f)}\; \left( b_{-r}^i\,b_{r}^j-
(i\leftrightarrow j)\right)\right)\;\;\; .
\cr & &
\eea
The anomalies
\be
\Delta_m^{(b)}=\Delta_\frac{m}{2}^{(f)} = 2A -1 + ( \frac{c^\perp}{12} - 1 )\,(m^2-1)
\ee
show that the complete $SO(d+1)$ algebra closes \textit{on-shell}, $p^+=p^-=p^0$, in
both sectors provided $\;c^\perp = 12\;$ and $\;A=\frac{1}{2}\;$.

\bigskip

\section{Analysis of the spectrum and supersymmetry}
\cleqn

In this section, we will analyze the perturbative spectrum of the $D4$-$\bar D4$ system
defined in Section $2$.
Therefore, we must consider open superstrings with a time-like NN coordinate $X^0$,
four coordinates along the branes resumed in two complex fields
$Z^{(1)} \equiv X^1 + i X^2, Z^{(2)}\equiv X^3 + iX^4$, and five DD coordinates
$X^i , i=5,\dots, 9$ orthogonal to the branes.
Each coordinate field is paired with fermionic partners $\psi^0$ (Majorana),
$\Psi^{(1)} ,\Psi^{(2)}$ (Dirac's) and $\psi^i,i=5,\dots,9\,$ (Majoranas) respectively.
Furthermore, we take from the start the magnetic fields on the
$D4$ and $\bar D4$ branes to be $(B_1= T_s\,\tan(\frac{\pi}{2}\nu^{(1)}) , B_2= T_s\,
\tan(\frac{\pi}{2}\nu^{(2)}))$ and $(-B_1 ,-B_2)$ respectively, with $0<|\nu^{(i)}|<1$,
but with no relation between the $B_i$ 's.

According to the precedent section, the theory  obtained after fixing the light-cone
like gauge in the $(09)$ directions should present $SO(5)$ invariance, since $d=4$
and, in fact, $c^\perp =\frac{3}{2}\,4 + 6 =12$.
This invariance reflect in the spectrum, and we will give some examples below.

From (\ref{energy}) (with $A=\frac{1}{2}$), (\ref{vir}), (\ref{ve}), (\ref{barnu}),
(\ref{emtms}) and (\ref{vems}), the energy operator in any sector reads
\bea
E^2&\equiv& (p^0)^2 = \frac{1}{\alpha'}\; \left( L_0^\perp - \frac{1}{2}\right)\cr &=&
\delta_{\nu_-^{(1)},0}\;\cos^2\varphi_0^{(1)}\; |\vec p^{(1)}|^2
+\delta_{\nu_-^{(2)},0}\;\cos^2\varphi_0^{(2)}\; |{\vec p}^{(2)}|^2 +
\frac{1}{\alpha'}\; \left( N^\perp + N_0\right)\cr
N_0 &=& \Delta_0^{(1)} +\Delta_0^{(2)} + \sum_{i=5}^8\,\Delta_0^i - \frac{1}{2}=
\delta\;(\bar\nu^{(1)} + \bar\nu^{(2)}-1)\;\;\;\;,
\eea
where $N^\perp = N^{(1)} + N^{(2)} +  \sum_{i=5}^8\,N^i$ is the total number operator
in the eight transverse dimensions.

There are four ``CP factors", in the language of \cite{sen1}, which label the states
in the spectrum, depending on where the ends of the (oriented) strings are fixed;
we denote them by $dd, \bar d\bar d, d\bar d$ and $\bar d d$.
The vacuum energy $N_0$ depends on the CP label and on whether we are in the NS sector
($\delta=\frac{1}{2}$) or in R sector ($\delta=0$).
Furthermore, the spectrum must be GSO projected, procedure which preserves the states
such that
\be
(-)^{ \sharp+ \nu^{(1)}_{-} -\bar\nu^{(1)}+ \nu^{(2)}_{-}-\bar\nu^{(2)} }\;
\left\{\begin{array}{c} 1\cr \gamma\end{array}\right.= + 1
\ee
in the NS and R sectors respectively.
Here, $\sharp$ stands for the number of fermionic oscillators and
$\gamma\equiv -4\, b_0^5\, b_0^6\, b_0^7\, b_0^8\,\,$ is the chirality operator of $spin(4)$,
$\,\{b_0^i ;b_0^j\} = \delta^{ij}\,$ being the Clifford algebra.
The justification for this rule is given in the next Section, equation (\ref{gsoopen}).

Let us look first at the lowest levels.
In the $dd$ sector the ``massless" (null energy) spectrum is the well-known vector multiplet
corresponding to a theory with $16$ supercharges, consisting of a
a five dimensional $U(1)$ vector field, five scalars and the corresponding fermions;
idem in the $\bar d\bar d$ sector.
The $d\bar d$ superstrings, on the other hand, break half of this supersymmetry, but the
states which survive GSO are different in the following cases
\bigskip

\noindent\underline{\bf Case $ b^{(1)} b^{(2)} >0$ }
\bea
|0>_{NS} &,& \alpha'\,E^2 = \frac{\textit{sign}\; b^{(1)} }{2}\;
(|\nu_-^{(1)}| + |\nu_-^{(2)}| -1)\cr
B^{(1)}_{\bar\nu^{(1)} -\frac{1}{2}}\; B^{(2)}_{\bar\nu^{(2)} -\frac{1}{2}}\,|0>_{NS}
&,& \alpha' \,E^2 =\frac{\textit{sign}\; b^{(1)} }{2}\;(1- |\nu_-^{(1)}| -
|\nu_-^{(2)}|)\cr
P_+|\alpha>  &,& \alpha'\,E^2 = 0\;\;\;,\;\;\;\alpha=1,\dots,4\label{massless++}
\eea
\noindent\underline{\bf Case $ b^{(1)} b^{(2)} < 0$ }
\bea
B^{(a)}_{\bar\nu^{(a)} -\frac{1}{2}}\,|0>_{NS} &,&
E^2 =\frac{\textit{sign}\; b^{(a)} }{2\alpha'}\;(1- |\nu_-^{(1)}| - |\nu_-^{(2)}|)\cr
P_-|\alpha>  &,& \alpha'\,E^2 = 0\;\;\;,\;\;\;\alpha=1,\dots,4\label{massless+-}
\eea
where the states in the last lines are in the R sector, $|\alpha>$ being the spinor
representation of $spin(4)$ algebra and $P_\pm\equiv \frac{1}{2}\,(1 \pm\,\gamma)$.
We see that, under the condition (\ref{susycons}), the potential tachyons disappear and the
NS sector contributes to the massless level with two complex fields (considering the
$\bar d d$ superstrings).
These, together with the fermions, fill a $4 + 4 = 8$ dimensional
representation of a superalgebra with eight supercharges, corresponding to a massless
hypermultiplet \cite{polcho2, wein3}.

Let us make a further step and write down the first two massive levels (in the inter-brane
sector) indicating, in the first (second) column, the NS (R) sector states.
\bigskip

\noindent\underline{\bf{Case $ b^{(1)} >0\;,\; b^{(2)} >0$}}
\bigskip
\begin{itemize}

\item $\alpha'\,E^2 = |\nu_-^{(1)}|= \bar\nu^{(1)} = 1 - \bar\nu^{(2)}$
\bea
\begin{array}{r}
A^{(1)}_{\bar\nu^{(1)}}{}^\dagger\,|0>_{NS}\cr
A^{(2)}_{\bar\nu^{(2)}-1}{}\,B^{(1)}_{\bar\nu^{(1)} -\frac{1}{2}}\; B^{(2)}_{\bar\nu^{(2)}
-\frac{1}{2}}|0>_{NS}\cr
\left(A^{(1)}_{\bar\nu^{(1)}}{}^\dagger\,B^{(1)}_{\bar\nu^{(1)} -\frac{1}{2}}\,
B^{(2)}_{\bar\nu^{(2)} -\frac{1}{2}} - A^{(2)}_{\bar\nu^{(2)}-1}\right)|0>_{NS}\cr
B^{(2)}_{\bar\nu^{(2)}-\frac{1}{2}}\, b^i_{-\frac{1}{2}}\,|0>_{NS}\;,\;
A^{(2)}_{\bar\nu^{(2)}-1}\,|0>_{NS}
\end{array}\;\;\;\;\;\;
\begin{array}{rl}
A^{(a)}_{\bar\nu^{(a)}}{}^\dagger\, P_+|\alpha>\cr
B^{(a)}_{\bar\nu^{(a)}}{}^\dagger\,P_-|\alpha>
\end{array}
\eea

\item $\alpha'\,E^2 = |\nu_-^{(2)}|= \bar\nu^{(2)} = 1 - \bar\nu^{(1)}$
\bea
\begin{array}{r}
A^{(2)}_{\bar\nu^{(2)}}{}^\dagger\,|0>_{NS}\cr
A^{(1)}_{\bar\nu^{(1)}-1}\,B^{(1)}_{\bar\nu^{(1)} -\frac{1}{2}}\,
B^{(2)}_{\bar\nu^{(2)} -\frac{1}{2}}|0>_{NS}\cr
\left(A^{(2)}_{\bar\nu^{(2)}}{}^\dagger\,B^{(1)}_{\bar\nu^{(1)}-\frac{1}{2}}\;
B^{(2)}_{\bar\nu^{(2)}-\frac{1}{2}} - A^{(1)}_{\bar\nu^{(1)}-1}\right)|0>_{NS}\cr
B^{(1)}_{\bar\nu^{(1)}-\frac{1}{2}}\,b^i_ {-\frac{1}{2}}\,|0>_{NS}\;\;\;,\;\;\;
A^{(1)}_{\bar\nu^{(1)}-1}\,|0>_{NS}
\end{array}\;\;\;\;\;\;
\begin{array}{r}
A^{(a)}_{\bar\nu^{(a)}-1}\, P_+|\alpha>\cr
B^{(a)}_{\bar\nu^{(a)} -1} P_-|\alpha>
\end{array}
\eea
\end{itemize}
\bigskip

\noindent\underline{\bf{Case $ b^{(1)} <0\;,\; b^{(2)} <0$}}
\bigskip

As the precedent one, with the exchanging of levels
$|\nu_-^{(1)}|\leftrightarrow |\nu_-^{(2)}|$.
\bigskip

\noindent\underline{\bf{Case $ b^{(1)} >0\;,\; b^{(2)} <0$}}
\bigskip

\begin{itemize}

\item $\alpha'\,E^2 = |\nu_-^{(1)}|= \bar\nu^{(1)} = \bar\nu^{(2)}$
\bea
\begin{array}{r}
A^{(1)}_{\bar\nu^{(1)}}{}^\dagger\,B^{(2)}_{\bar\nu^{(2)}-\frac{1}{2}}\,|0>_{NS}\cr
A^{(2)}_{\bar\nu^{(2)}}{}^\dagger\, B^{(1)}_{\bar\nu^{(1)}-\frac{1}{2}}\,|0>_{NS}\cr
\left(A^{(1)}_{\bar\nu^{(1)}}{}^\dagger\, B^{(1)}_{\bar\nu^{(1)}-\frac{1}{2}} -
A^{(2)}_{\bar\nu^{(2)}}{}^\dagger\, B^{(2)}_{\bar\nu^{(2)}-\frac{1}{2}}\right)|0>_{NS}\cr
b^i_{-\frac{1}{2}}\,|0>_{NS}\;\;,\;\;
\left(A^{(1)}_{\bar\nu^{(1)}}{}^\dagger\, B^{(1)}_{\bar\nu^{(1)}-\frac{1}{2}} +
A^{(2)}_{\bar\nu^{(2)}}{}^\dagger\, B^{(2)}_{\bar\nu^{(2)}-\frac{1}{2}}\right)|0>_{NS}\cr
\end{array}\;\;\;\;\;\;
\begin{array}{r}
A^{(a)}_{\bar\nu^{(a)}}{}^\dagger\, P_-|\alpha>\cr
B^{(a)}_{\bar\nu^{(a)}}{}^\dagger\, P_+|\alpha>
\end{array}
\eea

\item $\alpha'\,E^2 = |\nu_-^{(2)}|= 1 - \bar\nu^{(2)} = 1 - \bar\nu^{(1)}$
\bea
\begin{array}{r}
A^{(1)}_{\bar\nu^{(1)}-1}\,B^{(2)}_{\bar\nu^{(2)}-\frac{1}{2}}\,|0>_{NS}\cr
A^{(2)}_{\bar\nu^{(2)}-1}\,B^{(1)}_{\bar\nu^{(1)}-\frac{1}{2}}\,|0>_{NS}\cr
\left( A^{(1)}_{\bar\nu^{(1)}-1}\,B^{(1)}_{\bar\nu^{(1)}-\frac{1}{2}} -
A^{(2)}_{\bar\nu^{(2)}-1}\,B^{(2)}_{\bar\nu^{(2)}-\frac{1}{2}}\right)|0>_{NS}\cr
b^i_{-\frac{1}{2}}\,|0>_{NS}\;\;,\;\;
\left( A^{(1)}_{\bar\nu^{(1)}-1}\,B^{(1)}_{\bar\nu^{(1)}-\frac{1}{2}} +
A^{(2)}_{\bar\nu^{(2)}-1}\,B^{(2)}_{\bar\nu^{(2)}-\frac{1}{2}}\right)|0>_{NS}
\end{array}\;\;\;\;\;\;
\begin{array}{rcr}
A^{(a)}_{\bar\nu^{(a)}-1} P_-|\alpha>\cr
B^{(a)}_{\bar\nu^{(a)} -1} P_+|\alpha>
\end{array}
\eea
\end{itemize}
\bigskip

\noindent\underline{\bf{Case $ b^{(1)} >0\;,\; b^{(2)} <0$}}
\bigskip

As the precedent one, with the exchanging of levels
$|\nu_-^{(1)}|\leftrightarrow |\nu_-^{(2)}|$.
\bigskip

In each case and level, we have arranged the spectrum in such a way that the three NS states
in the first three lines are $SO(5)$ scalars, while the five states in the last line
form an $SO(5)$ vector; in the R sector, each value of $a=1,2$ labels a $Spin(5)$ Dirac spinor.
These assertions can be easily checked by applying (\ref{lcgen}) on the states and, of
course, they signal the $SO(5)$ invariance of the spectrum.
In any case, the states in each level expand a  $16 + 16 = 32$ dimensional representation of
a superalgebra with eight supercharges, corresponding to a massive (non BPS) supermultiplet
\cite{polcho2, wein3}.

\section{Vacuum amplitudes and boundary states}
\cleqn

\subsection{ The open string channel}

Let us consider a $Dp$-brane located at $X^i=y_0^i\,,\,i=p+1,\dots,D-1$ and  another
one (or a $\bar Dp$-brane) at $X^i=y_\pi^i\,,\,i=p+1,\dots,D-1$ in $D$-dimensional flat
space. By means of a straightforward generalization of (\ref{accionmixed}), (\ref{bcm})
for arbitrary constant field strengths $F_0{}^\mu{}_\nu = T_s\, f_0{}^\mu{}_\nu$ and
$F_\pi{}^\mu{}_\nu = T_s\, f_\pi{}^\mu{}_\nu$ living on their world-volumes along the
$\mu =0, 1, \dots,p$ directions, the b.c. for the coordinate fields of open strings
suspended between them result
\be
\partial_\sigma X^\mu(\tau,0) - f_0{}^\mu{}_\nu\;\partial_\tau X^\nu(\tau,0)
=\partial_\sigma X^\mu(\tau,\pi) - f_\pi{}^\mu{}_\nu\;\partial_\tau X^\nu(\tau,\pi)= 0
\;\;\;.\label{oxbc}
\ee
The {\it one-loop} interaction diagram is constructed by imposing conditions of
periodicity (P), $\eta_\pm =+1$, or anti-periodicity (AP), $\eta_\pm =-1$,
in the euclidean time variable $\tau^e \equiv i\tau \sim \tau^e + T$
\be
X^M(\tau,\sigma) \sim X^M(\tau -i T,\sigma)
\;\;\;\;,\;\;\;\; \psi_\pm^M(\tau,\sigma) \sim \eta_\pm\;\psi_\pm^M(\tau -i T,\sigma)
\;\;\; .
\label{openbc}
\ee
This is carried out by taking the traces in the Hilbert space, remembering that,
in the case of fermions and ghost system with P b.c., we must insert the spinor number
operator (written for convenience in pairs of indices, see (\ref{u1cur}))
\be
(-)^{F^{\Psi}} = \prod_{a=1}^5(-)^{-J_0^{\Psi^{(a)}}}\;\;\;\;.
\ee
The  $\lambda=2$ ghost fields $b$-$c$ and $\lambda =\frac{3}{2}$ superghost fields
$\beta$-$\gamma$ follow the b.c. of the (bosonic) reparameterization and (fermionic)
SUGRA transformations parameters respectively.
The insertion of the spinor number operators
\be
(-)^{F^{bc}} = (-)^{U_0^{bc}}\;\;\;\;,\;\;\;\;
(-)^{F^{\beta\gamma}} = (-)^{U_0^{\beta\gamma}}
\ee
must be carried out when P (AP) b.c. apply, due to the fermionic (bosonic) character of the
ghost (superghost) system; the definition of the $U_0$ charges is given in
(\ref{gnc}).

The connected part of the one loop amplitude is guessed from the Coleman-Weinberg formula
\be
{\it A}^{1-loop} \equiv \ln\;Z^{1-loop}\sim -\frac{1}{2}\;tr\;(-)^{\bf F}\ln\;G^{-1}
\ee
where $G^{-1}= p^2 + M^2 = \alpha'^{-1}\,L_0$ is the inverse (free) propagator, ${\bf F}$
is the \textit{space-time} fermion number and the traces are on the full Hilbert space.
Regulating as usual the logarithm, we define
\bea
{\cal A}^{open} &=& -\frac{1}{2}\;tr_{NS}\; \ln\frac{G^{-1}}{T_s}
+\frac{1}{2}\;tr_{R}\; \ln\frac{G^{-1}}{T_s} =
\int_0^\infty\;\frac{dt}{2\,t}\;\left(A_{NS}^{open}(it) + A_R^{open}(it)\right)\cr
A_{NS,R}^{open}(\tau)&=&  tr_{NS,R}\; q^{L_0}\;(-)^{F^{\Psi}+ F^{bc}}|_{q=e^{i2\pi\tau}}
= A_{open}^{(b)}(\tau)\; A_{open}^{(f)}(\tau)|_{NS,R}\;\;\; .\label{ampop}
\eea
In what follows, we focus on our system.
The bosonic contribution is
\bea
A_{open}^{(b)}(it) &\equiv& Z^{X^0}(\tau)\; \prod_{i=5}^9\, Z^{X^i}(\tau)\; Z^{Z^{(1)}}(\tau)\;
Z^{Z^{(2)}}(\tau)\; Z^{bc}(\tau)\cr
&=& i\, V_5\; e^{i\pi (1-\bar\nu^{(1)} -\bar\nu^{(2)})}\frac{16\,b^{(1)}\,b^{(2)}}
{(8\pi^2\alpha')^\frac{5}{2}}\;
\frac{e^{-T_s\, \Delta \vec y^2\,t\,}}{t^{\frac{1}{2}}\;\eta(it)^{4}}\;
\left(Z^{1-2\bar\nu^{(1)}}_1 (it)\;Z^{1-2\bar\nu^{(2)}}_1 (it)\right)^{-1}\;\;,\cr
& &\label{ampbop}
\eea
where we have used the formulae given in (\ref{pf}).
\footnote{
In $Z^{bc}$ the zero mode sector must be projected out \cite{polcho1}.
}

In the fermionic sector, we must impose the GSO condition which in the presence of non
trivial b.c., becomes a little bit subtle.
It is equivalent to inserting in the traces the projection operator
\be
P_{GSO} \equiv \frac{1}{2}\;\left( 1 - (-)^{\nu_-^{(1)}+\nu_-^{(2)}}\;(-)^{F^{\Psi}
+F^{\beta\gamma}}\right)\;\;\;\;.\label{gsoopen}
\ee
The logic for this definition relies in two facts,
\begin{itemize}
\item $P_{GSO}{}^2 = P_{GSO}$ must hold in the Hilbert space of the (perturbative) theory;

\item On physical grounds it should reduce to the well-known operator
\be
P_{GSO}\equiv \frac{1}{2}\;\left( 1 - (-)^{F^{\Psi}+F^{\beta\gamma}}\right)
\ee
if we turn off adiabatically the gauge fields (note that the definition is in terms of
$\nu_-^{(a)}$, \textit{not} of $\bar\nu^{(a)}$, see (\ref{barnu}); note also the sign
flipped w.r.t. a $Dp$-$Dp$ system \cite{sen1}).
\end{itemize}

Carrying out the computations we get
\bea
A_{open}^{(f)}(it) &=& A_{open}^{(f);GSO}(it)|_{NS} + A_{open}^{(f);GSO}(it)|_R\cr
A_{open}^{(f);GSO}(it)|_{NS} &\equiv & tr_{NS}\; \prod_{a=1}^5\; q^{L_0^{\Psi^{(a)}}-
\frac{1}{24}}\;(-)^{F^{\Psi} + \sum_{a=1}^5 q_0^{(a)}}\;  q^{L_0^{\beta\gamma}-
\frac{11}{24} }\;(-)^{[\pi_0]}\; P_{GSO}\cr
&=& e^{i\pi( \bar\nu^{(1)} +\bar\nu^{(2)} )}\;\frac{1}{2}\,\left( Z^0_1(it)^2\;
Z^{-2\nu_-^{(1)}}_1 (it)\; Z^{-2\nu_-^{(2)}}_1(it)\right.\cr
&+& \left.e^{-i\pi(\nu_-^{(1)}+\nu_-^{(2)})}\;
Z^0_0(it)^2\; Z^{-2\nu_-^{(1)}}_0(it)\; Z^{-2\nu_-^{(2)}}_0 (it)\right)\cr
A_{open}^{(f);GSO}(it)|_{R} &\equiv & tr_R\; \prod_{a=1}^5\; q^{L_0^{\Psi^{(a)}}-
\frac{1}{24}}\;
(-)^{F^{\Psi} + \sum_{a=1}^5 q_0^{(a)}}\;  q^{L_0^{\beta\gamma}- \frac{11}{24} }\;
(-)^{[\pi_0]}\; P_{GSO}\cr
&=&e^{i\pi( \bar\nu^{(1)} +\bar\nu^{(2)} )}\;\frac{1}{2}\,\left( -Z^1_1(it)^2\;
Z^{1-2\nu_-^{(1)}}_1 (it)\; Z^{1-2\nu_-^{(2)}}_1(it)\right.\cr
&-& \left.e^{-i\pi(\nu_-^{(1)}+\nu_-^{(2)})}\;
Z^1_0(it)^2\; Z^{1-2\nu_-^{(1)}}_0(it)\; Z^{1-2\nu_-^{(2)}}_0(it)\right)\;\;\;.\label{ampfop}
\eea
From ({\ref{ampop}}), ({\ref{ampbop}}), ({\ref{ampfop}}), we obtain the final result
\footnote{
The bosonic and fermionic amplitudes are  separately invariant under
$\nu_-^{(a)}\rightarrow -\nu_-^{(a)}$; it follows that the contribution from exchanging
the ends of the strings is taken into account just with the introduction of a factor of two,
as was made in (\ref{openamplit}).
}
\bea
& &{\cal A}^{open} = -i\, V_5\;\frac{32\,b_0^{(1)}\,b_0^{(2)}}
{(8\pi^2\alpha')^\frac{5}{2}}\;\;\int_0^\infty\;\frac{dt}{2\,t}\;
\frac{ e^{-T_s\, \Delta \vec y^2\,t\,}}{t^{\frac{1}{2}}\;\eta(it)^{4}}\;
\left(Z^{1-2\nu_-^{(1)}}_1(it)\;Z^{1-2\nu_-^{(2)}}_1(it)\right)^{-1}\cr
& &\frac{1}{2}\,\left( Z^0_1(it)^2\; Z^{-2\nu_-^{(1)}}_1 (it)\; Z^{-2\nu_-^{(2)}}_1(it)
+ e^{-i\pi(\nu_-^{(1)}+\nu_-^{(2)})}\;
Z^0_0(it)^2\; Z^{-2\nu_-^{(1)}}_0(it)\; Z^{-2\nu_-^{(2)}}_0 (it)\right.\cr
& &\left. -Z^1_1(it)^2\; Z^{1-2\nu_-^{(1)}}_1 (it)\; Z^{1-2\nu_-^{(2)}}_1(it)
- e^{-i\pi(\nu_-^{(1)}+\nu_-^{(2)})}\;
Z^1_0(it)^2\; Z^{1-2\nu_-^{(1)}}_0(it)\; Z^{1-2\nu_-^{(2)}}_0(it)\right)\;\;.\cr
& &\label{openamplit}
\eea

\subsection{ The closed string channel}

The open string channel description of the interaction between branes just given has a
dual closed string channel picture as exchange of \textit{closed} strings between the
so called ``boundary states" (b.s.) $|B>$, that represent each brane as a sort of condensate
of closed strings (see \cite{dvl} for a review in the covariant formalism, which we will
adopt; for a light-cone approach, see \cite{greengut}).
They are completely determined, up to normalization, by conditions that can be guessed
by considering the conformal map
\be
w\equiv \tau^e + i\sigma = i\sigma^+  \longrightarrow \hat w =  -i\,\frac{\pi}{T}\,w =
\hat\tau^e + i\hat\sigma = i\hat\sigma^+ \Longleftrightarrow
\hat\sigma ^\pm = \mp i\frac{\pi}{T}\, \sigma^\pm\;\;\; .
\ee
Clearly, the coordinate $\hat\sigma = - \frac{\pi}{T}\tau^e \sim \hat\sigma +\pi$ becomes
the (periodic) spatial coordinate of the closed string, while
$\hat\tau^e =\frac{\pi}{T}\sigma \in [0, \hat T = \frac{\pi^2}{T}]$.
This draws the {\it tree level} cylinder interaction diagram between a b.s. at $\hat\tau=0$
and a b.s. at $\hat\tau=\hat T$, dual to the one-loop open string diagram just computed.

Under this conformal map general $\lambda$-tensors transform as
\be
t^{(\lambda)}_\pm d^\lambda\sigma^\pm = \hat t^{(\lambda)}_\pm d^\lambda\hat\sigma^\pm
\;\;\;\Longleftrightarrow\;\;\;\hat t^{(\lambda)}_\pm =
\left(\mp\,\frac{i\,\pi}{T}\right)^{-\lambda}\; t^{(\lambda)}_\pm\;\;\;.\label{octr}
\ee
This relation is crucial in order to get the right b.s. definition.
While, in the $\hat\sigma$- coordinate, the P (or AP) conditions just give the usual
closed string expansions,
\footnote{
We follow the conventions of reference \cite{GSW1}.
}
the conditions on $\hat\tau^e$ (coming from the $\sigma$ conditions in the open string
picture) are interpreted as operator equations to be satisfied by the b.s. $ |B>$.
The amplitude between two branes is then given by a closed superstring matrix element
\bea
{\cal A}^{closed}&\equiv& <B'|\, G\;\delta\left( L_0 - \tilde L_0\right)\, |B> =
\frac{\alpha'}{4\,\pi}\;\int_{|z|<1}\; \frac{d^2z}{z\,\bar z}\;
<B'|\,z^{L_0}\; \bar z^{\tilde L_0}\;\, |B>\cr
&=& \int_0^\infty\; \frac{dt}{2\,t}\; \pi\;\alpha'\;t\; A^{closed}(it)\;\;\;\;,
\label{closedamplitud}
\eea
where $L_0$ and $\tilde L_0$ are the total zero mode Virasoro generators in the left and
right sectors and
$G \equiv (p^2 + M^2)^{-1} = \frac{\alpha'}{2}\,(L_0 +\tilde L_0 )^{-1}$
is the closed superstring propagator.
Moreover from the factorization of the b.s.
\be
A^{closed}(it) = A^X(z,\tilde z)\;A^{bc}(z,\tilde z) \;A^\psi(z,\tilde z)\;
A^{\beta\gamma}(z,\tilde z)|_{z=\bar z =e^{-\pi\,t}} = A_{closed}^{(b)}(it)\;
A_{closed}^{(f)}(it)\;\;.\label{closedamplit}
\ee
We remark that the ``bra" b.s. must be defined in such a way that for any ``ket" $|\psi>\;$,
$<B|\psi> \equiv \left(|B>;\,|\psi>\right)\;$ for a hermitian scalar product
$<\xi|\psi>^*=<\psi|\xi>$ which respects the hermiticity conditions of the fields under
consideration.

We resume below the b.s. at $\tau=0$ (we throw away the subindex ``$0$")
\footnote{
It is evident, from (\ref{closedamplitud}) and the fact that the evolution operator is
$H=L_0+\tilde L_0$, that the amplitude is independent of the value of $\tau$ at
which we compute it.
}
as well as the amplitudes $A(z,\tilde z)'s$ for each field.
\bigskip

\noindent{\underline{Scalar sector boundary state}}
\bigskip

The b.c. (\ref{oxbc}) for the scalar periodic fields $X^M$ become, in the
closed channel,
\bea
\left( \partial_{\hat\sigma^+} \hat X^\mu (0,\hat\sigma)  + S^\mu{}_\nu\;
\partial_{\hat\sigma^-} \hat X^\nu (0,\hat\sigma) \right)\;|B^X> &=&0\;\;\;\;\;\;\;\;\;\;
\;\;\;\;\;,\;\;\;
\mu,\nu=0,1,\dots p\cr
\hat X^i (0,\hat\sigma) \;|B^X> &=& y_0^i\;|B^X>\;\;\;,\;\;\; i=p+1,\dots,p + d_\perp\;\;,\cr
& &\label{obbc}
\eea
where $S \equiv (d^-){}^{-1}\,d^+\,,\,d^\pm = 1 \pm f\,$ and $d_\perp = D-1-p$.
Equivalently, in terms of modes, these conditions translate to
\vfill\eject
\bea
p^\mu\,|B^X> &=& 0\cr
x^i\,|B^X> &=& y_0^i\,|B^X>\cr
\left( \alpha_m^M + M^M{}_N\;\tilde\alpha_{-m}^N\right)\,|B^X> &=& 0\;\;\;,\;\;\;m\neq0
\;\;\;\;,
\label{obbcm}
\eea
where $\,M \equiv \left( \matrix{ S^{-1}& 0\cr 0 & -1_{d_\perp} }\right)\,$.
The solution for the b.s. defined by (\ref{obbc}) is
\bea
|B^X> &=& N_B\;\prod_{m=1}^\infty\
e^{-\frac{1}{m}\alpha_{-m}^M\,M_{MN}\,\tilde\alpha_{-m}^N}\,|B^X>_0\cr
|B^X>_0 &=& |p^\mu=0;x^i=y_0^i>\otimes |0>\;\;\;\;,
\eea
where we have attached to it the normalization constant $N_B$.
The amplitude is computed by using the usual pairing defined by the hermiticity conditions
$\alpha_m^M{}^\dagger = \alpha_{-m}^M\;$, $\;\tilde\alpha_m^M{}^\dagger =
\tilde\alpha_{-m}^M\;$ for the left and right oscillators and
$<0|0> \equiv 1\;,\; <p'|p> \equiv \delta^D (p-p')$
in the oscillator and zero mode sectors respectively.
We get
\bea
& &A^X(z,\tilde z)
\equiv <B'{}^X |\,z^{L_0^X - \frac{10}{24}}\;\bar z^{\tilde L_0^X -\frac{10}{24}}\;
|B^X> \cr
&=& N'_B{}^*\; N_B\;V_{p+1}\; \left(\frac{T_s}{\log|z|^{-1}}\right)^{\frac{d_\perp}{2}}\;
\frac{e^{-\frac{\pi T_s \Delta y^2}{\log|z|^{-1}}}}{|z|^{\frac{5}{6}}}\; \prod_{m=1}^{\infty}\;
\det\left(1_D - |z|^{2m}\; M^{-1}\;M'\right)^{-1}\;\;\;,\cr
& &
\eea
where $\Delta y = \sqrt{(  \vec y'_0 - \vec y_0)^2}$ is the separation between branes
(to be taken to zero).
\bigskip

\noindent{\underline{$b$-$c$ ghost boundary state}}
\bigskip

It is straightforward to see, with the help of (\ref{octr}), that the open string b.c.
(\ref{obcbc}) of a $\lambda=2$ anticommuting, periodic $b$-$c$ system translate, in the
closed channel, into the following conditions for the b.s. $|B^{bc}>$
\bea
\left(\hat b (0,\hat\sigma) - \hat{\tilde b} (0,\hat\sigma)\right)\,|B^{bc}> &=& 0
\;\;\;\; \longleftrightarrow \;\;\;\;
\left(\hat b_m - \hat{\tilde b}_{-m}\right)\,|B^{bc}> = 0\cr
\left(\hat c (0,\hat\sigma) + \hat{\tilde c} (0,\hat\sigma)\right)\,|B^{bc}> &=& 0
\;\;\;\; \longleftrightarrow \;\;\;\;\;
\left(\hat c_m + \hat{\tilde c}_{-m}\right)\,|B^{bc}> = 0\;\;,\label{bcbsd}
\eea
for any $m\in\bz$.
The b.s. that solves (\ref{bcbsd}) is
\footnote{
The zero mode ghost sector is realized by defining the four states
\be
|s\tilde s>\equiv |s>\otimes|\tilde s>\;\;\;,\;\;\; s, \tilde s = +,-\;\;\;\;,
\ee
on which the zero mode operators act as
\be
c_0 = \sigma_+\otimes 1_2\;\;\;,\;\;\;
b_0 = \sigma_-\otimes 1_2\;\;\;,\;\;\;
\tilde c_0 = \sigma_3\otimes \sigma_+\;\;\;,\;\;\;
\tilde b_0 = \sigma_3\otimes \sigma_-\;\;\;\;,
\ee
where $\sigma_\pm = \sigma_1 \pm i\sigma_2$ and $\sigma_i$ are the Pauli matrices.
As usual, for the other operators
\be
b_m |s\tilde s> = \tilde b_m |s\tilde s> = c_{-m} |s\tilde s> = \tilde c_{-m} |s\tilde s>
= 0 \;\;\;\;,\;\;\;\ m=1,2,\dots\;\;\;\;.
\ee
}
\bea
|B^{bc}> &=& \prod_{m=1}^\infty\
e^{c_{-m}\,\tilde b _{-m} - b_{-m}\,\tilde c _{-m}}\;|B^{bc}>_0\cr
|B^{bc}>_0 &=&  \frac{1}{\sqrt 2}\,\left( |+-> - |-+>\right)\;\;\;\;.
\eea
The hermiticity conditions,
$\;b_m{}^\dagger = b_{-m}\;,\; c_m{}^\dagger = c_{-m}\;,\;\tilde b_m{}^\dagger =
\tilde b_{-m}\;,\;\tilde c_m{}^\dagger = \tilde c_{-m}\;$,
impose that the only non-zero pairings are
\be
<+-|-+> = -<-+|+-> = i\;\;\;\;.
\ee
In particular, $\;\;{}_0<B^{bc}|B^{bc}>_0 = \frac{1}{2}\,(-i + i)=0\;$; however, the
amplitude is
\be
A^{bc}(z,\bar z)
\equiv <B{}^{bc} |(b_0+\tilde b_0)\;(c_0 -\tilde c_0)\;\,z^{L_0^{bc}+ \frac{26}{24}}\;
\bar z^{\tilde L_0^{bc}+ \frac{26}{24}}\; |B^{bc}>
= i\;|z|^\frac{1}{6}\;\prod_{m=1}^{\infty}\;\left(1 - |z|^{2m}\right)^2\;\;,
\ee
where the zero mode insertions coming from the measure (taken into account in (\ref{ampbop}))
are translated according to (\ref{octr}), (\ref{bcinv}) \cite{polcho1}.
\bigskip

\noindent{\underline{Fermionic boundary state}}
\bigskip

We will heavily rely on superconformal invariance to carry out the analysis.
Given $D=10$ Majorana fermions, the open string boundary term (\ref{bt1})
\be
i\,\frac{T_s}{2}\int\;d\tau\; \eta_{MM}\;(\delta\psi_+^M\;\psi_+^N -
\delta\psi_-^M\;\psi_-^N)|_{\sigma=0}^{\sigma=\pi}
\ee
cancels when the following b.c. hold
\be
\psi_-^M(0,\sigma) = \Lambda_0{}^M{}_N\; \psi_+^N (0,\sigma) \;\;\;\;,\;\;\;\;
\psi_-^M(\pi,\sigma) = \Lambda_\pi{}^M{}_N\; \psi_+^N (\pi,\sigma)\;\;\;,\label{ofbcD}
\ee
where $\Lambda_0$ and $\Lambda_\pi$ are arbitrary $O(1,9)$ matrices.
However, once we embed this theory into a superstring one by adding $D=10$
bosonic partners with b.c. (\ref{oxbc}), compatibility with the
superconformal transformations (\ref{SUSY}) fix them almost uniquely to be
\be
\Lambda_0 = \eta_0\; M_0\;\;\;\;,\;\;\;\; \Lambda_\pi = \eta_\pi\; M_\pi \;\;\;,\;\;\;
\eta_0{}^2 =\eta_\pi{}^2 =1\;\;\;\;,\label{lambdaM}
\ee
together with the periodicity of the SUSY parameter
\be
\epsilon^+(\tau) = -\eta_0\; \epsilon^-(\tau) = \eta_0\,\eta_\pi\;\epsilon^+(\tau + 2\,\pi)
\;\;\longleftrightarrow\;\;
\epsilon^+(\sigma^+)|_{\sigma=0,\pi} = -\eta_{0,\pi}\;\epsilon^-(\sigma^-)|_{\sigma=0,\pi}
\;\;.
\ee
It is worth to note that the matrices $M_{0,\pi}$ (as defined below (\ref{obbcm}))
belong, in fact, to $SO(1,p)\subset O(1,9)$, as it can be readily checked.

Around the loop, the fermions can be P or AP, according to (\ref{openbc});
from (\ref{octr}) (with $\lambda=\frac{1}{2}$) they become, in the closed channel,
\be
\hat\psi_\pm^M(\hat\tau,\hat\sigma + \pi) = \eta_\pm\; \hat\psi_\pm^M(\hat\tau,\hat\sigma)
\;\;\;,\;\;\;\eta_\pm{}^2=1\;\;\;\;,
\ee
giving rise to the well-known four sectors of the closed string; R-R if
$\eta_+=\eta_-=+1$, NS-NS if $\eta_+=\eta_-=-1$, NS-R if $\eta_+=-\eta_-=+1$, and
R-NS if $\eta_+=-\eta_-=-1$.
Then, from (\ref{octr}), (\ref{ofbcD}) and (\ref{lambdaM}) (with $\eta_0\equiv\eta$),
we get the defining relations for the fermionic b.s.
\be
\left( \hat\psi_-^M(0,\hat\sigma) - i\,\eta\; M^M{}_N\;\hat\psi_+^N(0,\hat\sigma)\right)
\;|B^\psi;\eta> = 0\;\;\;\;.\label{fbsc}
\ee
It is easy to see that such a b.s. only exists in the $NS$-$NS$ and $R$-$R$ sectors;
in what follows, we will label them by $\delta=\frac{1}{2}$ and $\delta=0$ respectively.
Furthermore, at first sight there are two states in each one of the two sectors labelled
by $\eta=\pm 1$; however, the GSO projection to be considered below will allow for a linear
combination of them to survive.
In terms of modes, (\ref{fbsc}) is equivalent to
\be
\left( b_m^M - i\,\eta\; M^M{}_N\;\tilde b_{-m}^N  \right)\;|B^\psi;\eta> = 0
\;\;\;\;,\;\;\;\; \forall\; m\in\bz_\delta\;\;\;\;.
\label{fbsdef}
\ee
The solution is
\footnote{
The state $|0>$ is the usual vacuum in the NS-NS sector, defined by
\be
b_m^M \, |0> = \tilde b_m^M |0> =0\;\;\;,\;\;\;
m\;\in\;\bz_\frac{1}{2}^+\;\;\;\;,
\ee
while $|\Lambda\tilde\Lambda>\equiv |\Lambda>\otimes|\tilde\Lambda>$
is the R-R vacuum obeying
\bea
b_m^M \,|\Lambda\tilde\Lambda> &=& \tilde b_m^M |\Lambda\tilde\Lambda> =0
\;\;\;,\;\;\; m\;\in\;\bz^+\cr
b_0^M \,|\Lambda\tilde\Lambda> &=& \frac{1}{\sqrt2}\;
\Gamma^M{}^\Omega{}_\Lambda\;|\Omega\tilde\Lambda>\;\;\;\;;\;\;\;\;
\tilde b _0^M \,|\Lambda\tilde\Lambda> = \frac{1}{\sqrt2}\;
\Gamma_{11}{}^{\Omega}{}_{\Lambda}\;
\Gamma^M{}^{\tilde\Omega}{}_{\tilde\Lambda}\;|\Omega\tilde\Omega>\;\;\;.
\eea
}
\bea
|B^\psi;\eta_> &=& \prod_{m\in\bz_\delta^+}^\infty\;\e^{i\,\eta\,
 b_{-m}^M \, M_{MN}\,\tilde b_{-m}^N}\;|B^\psi;\eta>_0\cr
|B^\psi;\eta >_0 &=&\left\{\begin{array}{ll}|0> &,\;\;\;\; \delta=\frac{1}{2}\cr
|N>= N^{\Lambda\bar\Lambda}\;|\Lambda\tilde\Lambda> &,\;\;\;\; \delta=0
\end{array}\right.\;\;\;\;,\label{frrbs}
\eea
where, from (\ref{fbsdef}) with $m=0$, it follows that the matrix $N$ must satisfy
\be
\Gamma^M\; N = i\,\eta\; M^M{}_N\;\Gamma_{11}\; N\; \Gamma^N{}^t\;\;\;\;.
\ee
The solution is unique up to normalization; if
\bea
A_\pm\;\Gamma^M\; A_\pm{}^{-1} &=& \pm \Gamma^M{}^t\;\;\;\;\;\;\;\;\;\;\;\Longrightarrow\;\;\;
A_- = A_+\,\Gamma_{11}= i\,\Gamma_0\cr
U^{-1}\;\Gamma^M\; U &=& S^{-1}{}^M{}_N\;\Gamma^N\;\;\;\Longrightarrow \;\;\;\;\;U = S(S^{-1})
\;\;\;,
\eea
we can take it to be
\bea
N &=&  U\; \Gamma^{01\dots p}\; \gamma_\pm\; A_\pm{}^{-1} = N^*\;\Gamma_{11}\cr
\gamma_\pm &\equiv& \frac{1\pm i\,\eta\;\Gamma_{11}}{1\pm i\,\eta}=
\gamma_\mp\;\Gamma_{11}\;\;\;\;,
\eea
where $S(S^{-1})$ is the spinor representation of the element $\;S^{-1}\in SO(1,p)$.
Explicitly for the system under consideration,
\be
\;S(S_{0,\pi})=
S\left( \left(\begin{array}{cccc}1&0&0&0\cr 0& e^{i\,\pi\,\nu_{0,\pi}^{(1)}\,\sigma_2}&0&0\cr
0& 0& e^{i\,\pi\,\nu_{0,\pi}^{(2)}\,\sigma_2}&0\cr0&0&0&1_5\end{array}\right)\right) =
e^{\frac{\pi}{2}\,\nu_{0,\pi}^{(1)}\,\Gamma_{12}+
\frac{\pi}{2}\,\nu_{0,\pi}^{(2)}\,\Gamma_{34}}\;\;\;\;.
\ee
The hermiticity conditions
$\;\psi_m^M{}^\dagger = \psi_{-m}^M\;,\;\tilde\psi_m^M{}^\dagger = \tilde\psi_{-m}^M\;$,
for any $m\in\bz$ fix, almost completely (up to normalization), the scalar product to be
\bea
<0|0> &=& 1 \;\;\;\;\;\;\;\;\;\;\;\;\;\;\;,\;\;\;\; \delta=\frac{1}{2}\cr
<\Lambda\tilde\Lambda|\Omega\tilde\Omega> &=&
g_{\Lambda\Omega}\;\tilde g_{\tilde\Lambda\tilde\Omega}\;\;\;\;\;,\;\;\;\;\delta
=0\;\;\;\;,
\eea
where, in the R-R sector, the $m=0$ conditions
$\;\psi_0^M{}^\dagger = \psi_0^M\,,\,\tilde\psi_0^M{}^\dagger = \tilde\psi_0^M\;$, impose
\be
\left.\begin{array}{rcl} g\;\Gamma^M\; g^{-1} &=& \Gamma^M{}^\dagger\cr
g\;\Gamma_{11} &=& \pm \Gamma_{11}\; g\cr
\tilde g\;\Gamma^M\; \tilde g ^{-1} &=& \pm \Gamma^M{}^\dagger
\end{array}\right\}\Longrightarrow
\left\{\begin{array}{rcccccl}
g &=& -i\, C_+ &=&  \Gamma_0\;\Gamma_{11} &=& g^\dagger\cr
\tilde g &=& C_- &=& i\,\Gamma_0 &=& \tilde g^\dagger\end{array}\right.\;\;\;\;.
\label{sprr}
\ee
With these definitions we get
\bea
A^{\psi}(z,\tilde z;\eta,\eta') &\equiv& <B'^\psi;\eta'|\,z^{L_0^\psi -\frac{5}{24}}\;
\bar z^{\tilde L_0^\psi -\frac{5}{24}}\;\,
|B^\psi ;\eta>\cr
&=& |z|^{\frac{5}{6}(1-3\delta)}\; \prod_{m=1}^\infty\;\det
\left( 1_{10} + \eta\;\eta'\;|z|^{2(m-\delta)}\;
M\;M'^{-1}\;  \right)\; A_0^\psi\cr
A_0^\psi &\equiv& {}_0<B'_\psi;\eta'|B_\psi;\eta >_0 =
\left\{\begin{array}{ll}1 \;\;&,\;\;\;\; \delta=\frac{1}{2}\cr
-32\; \eta\; \cos \pi\nu_-^{(1)}\;\cos \pi\nu_-^{(2)}\;\delta_{\eta\eta',1}
\;\;&,\;\;\;\;\delta =0\end{array}\right.\;.\cr
& &
\eea

\noindent{\underline{$\beta$-$\gamma$ super-ghost boundary state}}
\bigskip

According to (\ref{obcbc}) the open string b.c. of a $\lambda=\frac{3}{2}$ commuting,
$\beta$-$\gamma$ system are
\vfill\eject
\bea
\tilde\beta (\tau,0) &=& \eta_0\; \beta(\tau,0) \;\;\;\;;\;\;\;\;\;
\;\;\;\;\;\;\tilde\beta(\tau,\pi) = \eta_\pi\; \beta (\tau,\pi)\cr
\beta(\tau-iT,\sigma) &\sim& \eta_+\;\beta(\tau,\sigma)\;\;\;;\;\;\;
\tilde\beta(\tau-iT,\sigma) \sim \eta_-\;\tilde\beta(\tau,\sigma)\;\;\;\;,\label{betagamaopbc}
\eea
and analogously for $\gamma, \tilde \gamma$.
The phases $\eta_0 \equiv \eta, \eta_\pi$ and $\eta_+ , \eta_-$ must be identified with
those introduced in (\ref{openbc}) and (\ref{lambdaM}) respectively, because they follow
the superconformal transformation parameter modding.
With the help of (\ref{octr}) the b.c. (\ref{betagamaopbc}) translate, in the closed
channel, into the following defining conditions for the b.s. $|B^{\beta\gamma};\eta>$
\bea
\left(\hat\beta (0,\hat\sigma) + i\;\eta\; \hat{\tilde\beta} (0,\hat\sigma)\right)\,
|B^{\beta\gamma};\eta>&=& 0\;\;\;\;\;\longleftrightarrow\;\;\;\;\;
\left(\hat \beta_m + i\;\eta\;\hat{\tilde \beta}_{-m}\right)\,|B^{\beta\gamma};\eta> = 0\cr
\left(\hat\gamma (0,\hat\sigma)+ i\;\eta\; \hat{\tilde\gamma} (0,\hat\sigma)\right)\,
|B^{\beta\gamma};\eta>&=& 0\;\;\;\;\;\longleftrightarrow\;\;\;\;\;
\left(\hat\gamma_m + i\;\eta\;\hat{\tilde\gamma}_{-m}\right)\,|B^{\beta\gamma};\eta> = 0
\;\;\;,\cr
& &\label{gbggbsd}
\eea
for any $m\in\bz_\delta$.
We will restrict ourselves to $\;\tilde \pi_0 = 1-\pi_0\;$, the ``soaking up" anomaly
condition, in which case (\ref{gbggbsd}) is solved by
\be
|B^{\beta\gamma};\eta>_{\pi_0,\tilde \pi_0} = \prod_{m\geq \pi_0}\;
e^{i\,\eta\,\gamma_{-m}\,\tilde \beta _{-m}}\;\prod_{m\geq \tilde \pi_0}\;
e^{-i\,\eta\, \beta_{-m}\,\tilde\gamma _{-m}}\;|0>_{\pi_0,\tilde \pi_0}
\ee
where $|0>_{\pi_0,\tilde \pi_0}$ is defined at left and right as in $(\ref{ghostvacuum})$.
However, there is a very important difference with the anticommuting ghosts;
in that case, in view of (\ref{identifications}), we took, with no loss of generality
$p_0 =\tilde p_0 = 0$.
But, due to the commuting character of the $\beta$-$\gamma$ system, there exists no such
identification, i.e., each pair $(\pi_0,\tilde \pi_0)$ defines a {\cal different}
representation of the superghost algebra, the so called ``pictures", denoted commonly
as $(-\frac{1}{2}-\pi_0, -\frac{1}{2}-\tilde \pi_0)$.
Therefore, the vacuum, as well as the physical (BRST invariant) operators, must be referred
to a particular picture, the different pictures being related by the ``picture changing
operation" of FMS \cite{fms}.
We will not dwell into details about these facts, but just restrict ourselves to consider
the tools necessary to define and compute the amplitudes.
The hermiticity properties
\be
\beta_m{}^\dagger = -\beta_{-m}\;,\; \gamma_m{}^\dagger = \gamma_{-m}\;,\;
\tilde\beta_m{}^\dagger = -\tilde\beta_{-m}\;,\;\tilde\gamma_m{}^\dagger =\tilde\gamma_{-m}
\ee
determine the scalar product, to be defined just by imposing
\be
{}_{\pi'_0,\tilde \pi'_0}~<0|0>_{\pi_0,\tilde \pi_0} \equiv \delta_{\pi'_0, 1- \pi_0}\;
\delta_{\tilde \pi'_0,1-\tilde \pi_0}\;\;\;\;.
\ee
We get for the amplitude in the $(-\frac{1}{2}-\pi_0, -\frac{3}{2}+\pi_0)$ picture
\bea
A^{\beta\gamma}(z,\tilde z;\eta,\eta') &\equiv& {}_{1-\pi_0,\pi_0}<B{}^{\beta\gamma};\eta'|\,
z^{L_0^{\beta\gamma}- \frac{11}{24}}\;\bar z^{\tilde L_0^{\beta\gamma}- \frac{11}{24}}\;\,
|B^{\beta\gamma};\eta>_{\pi_0, 1-\pi_0}\cr
&=& |z|^{\frac{1}{6}(3\delta -1)}\;
\prod_{m=1}^\infty\; \left( 1 + \eta\,\eta'\; |z|^{2(m-\delta)}\right)^{-2}\;
\left\{\begin{array}{ll} (1 + \eta\,\eta')^{-1}   &,\;\;\delta=0\cr
(\eta\,\eta')^{\pi_0-\frac{1}{2}} &,\;\;\delta=\frac{1}{2}\end{array}\right.\cr
&=& (\eta\,\eta')^{\pi_0 -\frac{1}{2}} \;\left(Z^{1-2\delta}_{2b}(it)|_{e^{i2\pi b}=
\eta\eta', |z|=e^{-\pi t}}\right)^{-1}\;\;\;\;.
\eea

Now, we are ready to construct the amplitude in the closed channel.
According to (\ref{closedamplit}), the bosonic sector gives
\bea
A^{(b)}_{closed}(it) &\equiv& A^X(z,\tilde z)\;A^{bc}(z,\tilde z)|_{z=\bar z =e^{-\pi\,t}}\cr
&=& i\,N_{B'}{}^*\; N_B\;V_5\; \frac{4\,\sin\pi\bar\nu^{(1)}\,\sin\pi\bar\nu^{(2)}}
{(2\pi^2\alpha')^\frac{5}{2}}\; \frac{e^{-\frac{T_s\,\Delta y^2}{t}}}{t^\frac{5}{2}\,
\eta(it)^4}\; \left(Z^1_{1+2\bar\nu^{(1)}}(it)\; Z^1_{1+2\bar\nu^{(2)}}(it)\right)^{-1}\;.\cr
& &\label{ampbclos}
\eea
In the fermionic sector, we must sum both contributions from the NS-NS and R-R sectors,
this last one with a ``minus" sign coming from the charge of the anti D-brane.
Furthermore, we must project GSO in each one, i.e. we must insert the operator \cite{polcho2}
\be
P_{GSO} \equiv \frac{1+ (-)^F}{2}\;\frac{1+ (-)^{(p+1)(1-2\delta)}\; (-)^{\tilde F}}{2}
\ee
in the computation of the traces, where $F= F^\Psi + F^{\beta\gamma}$ and
$\tilde F= \tilde F^\Psi + \tilde F^{\beta\gamma}$ are the left and right spinor number
operators and $p$ is even (odd) in the type IIA (IIB) theory.
Then
\footnote{
8The definition of such amplitudes is equivalent to considering the GSO-projected b.s.
\cite{yost}
\be
|B>_{GSO}\equiv P_{GSO}|B;+> = \frac{1}{2}\; \left( |B;+> +
(-)^{[\pi_0]+ 1 + p(1-2\delta)}|B;->\right)
\ee
}
\bea
A^{(f)}_{closed}(it) &\equiv& A_{NS-NS}^{(f);GSO}(it) - A_{R-R}^{(f);GSO}(it)\cr
A_{NS-NS}^{(f);GSO}(it) &\equiv& A^{\psi;GSO}( z,\tilde z;\eta,\eta' )\;
A^{\beta\gamma;GSO}(z,\tilde z ;\eta ,\eta')\; |_{\eta=\eta'=1}\cr
&=&\frac{1}{2}\,\left( Z^0_0(it)^2\; Z^0_{2\nu_-^{(1)}} (it)\; Z^0_{2\nu_-^{(2)}}(it)
- Z^0_1(it)^2\; Z^0_{1+2\nu_-^{(1)}}(it)\; Z^0_{1+2\nu_-^{(2)}}(it)\right)\cr
A_{R-R}^{(f);GSO}(it) &\equiv& A^{\psi;GSO}( z,\tilde z;\eta,\eta')\;
A^{\beta\gamma;GSO}(z,\tilde z;\eta ,\eta')\; |_{\eta=\eta'=1}\cr
&=& -\frac{1}{2}\; Z^1_0(it)^2\; Z^1_{2\nu_-^{(1)}}(it)\;
Z^1_{2\nu_-^{(2)}}(it)\;\;\;\;.
\label{ampfclos}
\eea
It follows that, if we normalize the b.s. according to
\be
N_B \equiv \left(\frac{(1 + {b^{(1)}}^2 )\;(1 + {b^{(2)}}^2 )}{16\,\pi\,\alpha'}
\right)^\frac{1}{2}\;\;\;\;,
\ee
from (\ref{ampbop}), (\ref{ampfop}), (\ref{ampbclos}), (\ref{ampfclos})
and using the modular properties (\ref{modtransf}), the following identities
hold
\bea
A^{(b)}_{open}(it^{-1}) &=&
sign(b^{(1)}\,b^{(2)})\;\pi\,\alpha'\,t\;A^{(b)}_{closed}(it)\cr
A^{(f)}_{open}(it^{-1}) &=&
sign(b^{(1)}\,b^{(2)})\;A^{(f)}_{closed}(it)\;\;\;\;.
\eea
It is then straightforward to prove that (\ref{closedamplitud}) coincides exactly with
(\ref{openamplit}).
\bigskip

\subsection{No force conditions and supersymmetry}

To write the amplitude (\ref{openamplit}) just obtained from both channels in a convenient
way, we can use the \textit{second addition theorem} for theta functions \cite{fay}
which states that the following quartic identity
\be
\prod_{i=1}^4\; \vartheta\left[\matrix{ a_i\cr b_i}\right](\nu_i;\tau)
= \frac{1}{2}\; \sum_{s_1,s_2=0,\frac{1}{2}}\;e^{-i4\pi a_1 s_2 }\;
\prod_{i=1}^4\; \vartheta\left[\matrix{ m_i + s_1\cr n_i +s_2}\right](\epsilon_i;\tau)
\ee
holds, where
\bea
\left(\matrix{\nu_1 \cr \vdots\cr \nu_4}\right)
&=&\frac{1}{2}\;\left(\matrix{1&1&1&1\cr1&1&-1&-1\cr1&-1&1&-1\cr1&-1&-1&1\cr}\right)\;
\; \left(\matrix{\epsilon_1 \cr \vdots\cr \epsilon_4}\right)\cr
\left(\matrix{ \left[\matrix{a_1 \cr b_1}\right]
\cr \vdots\cr \left[\matrix{a_4 \cr b_4}\right] }\right)
&=&\frac{1}{2}\;\left(\matrix{1&1&1&1\cr1&1&-1&-1\cr1&-1&1&-1\cr1&-1&-1&1\cr}\right)
\otimes 1_2\left(\matrix{ \left[\matrix{m_1 \cr n_1}\right]
\cr \vdots\cr \left[\matrix{m_4 \cr n_4}\right] }\right)\;\;\;\;.
\eea
The famous ``abstruse" identity is a special case of it.
By choosing the arguments $\epsilon_i = 0$ and the spin structures
$m_1 = -\nu_-^{(1)}\;,\; n_1 = 1\; ,\; m_2= -\nu_-^{(2)}\;$ and the other zero,
we get
\be
\frac{1}{2}\,(\dots)_{(\ref{openamplit})} = e^{ i\pi( \nu_-^{(1)} -\nu_-^{(2)} ) }\;
\left( Z^{-\nu_-^{(1)} -\nu_-^{(2)} }_1(it)\; Z^{-\nu_-^{(1)} +\nu_-^{(2)}
}_1(it)\right)^2\;\;\;\;.
\ee
By virtue of the identity $Z^1_1(it)\equiv 0$, this equation shows that the
$D4$-$\bar D 4$ amplitude is identically zero iff
\be
|\nu_-^{(1)}| + |\nu_-^{(2)}| =1 \;\;\;\longleftrightarrow\;\;\; |b^{(1)}\;b^{(2)}|
=1\;\;\;\;,
\ee
which is exactly the condition (\ref{susycons}) founded for SUSY to hold!

\section{Conclusions and perspectives}
\cleqn

In this paper we have studied  a $D4$-$\bar D 4$ system with non zero magnetic fields in the
world-volumes of the branes in the weak coupling regime, and we have shown that it becomes
stable and supersymmetric for values of the gauge fields that coincide with those
conjectured from the weak coupling, low energy approach defined by the
Dirac-Born-Infeld world-volume effective field theory on a brane.
We think these results provide further evidence about the existence of the system.
In particular, a solution of $D=10$ SUGRA IIA describing the long distance fields such
brane-antibrane system supports should exist, maybe as a certain limit of a more
general solution, a kind of ``four dimensional supertube",
much as it happens with the $D2$-$\bar D 2$ and the supertube.
Finding such solution is certainly a very interesting open problem.

Moreover, it is natural to ask for the effective five dimensional world-volume
field theory of the system.
A $U(1)$ gauge field $A$ ($\bar A$) and five scalars live on the brane
(anti-brane), that together with the eight fermionic degrees of freedom fill a
vector multiplet of the $D=5$, $SUSY_1$ algebra (remember that the system preserves $8$
supercharges).
Furthermore, the two ``ex-tachyons" in (\ref{massless++}), (\ref{massless+-}) are charged
under $A^-\equiv A - \bar A\,$ \cite{schwarz}; then, the spectrum analysis of Section $4$
reveals the existence of a hypermultiplet of such algebra, coming from the inter-branes
excitations and charged under $A^-$.
Finally, the commutators among the coordinate fields along the longitudinal directions
of the brane-antibrane system (see Appendix B)
\footnote{
In its evaluation the following identities valid for $\nu\notin\bz$ are required
(see page $46$ in \cite{rusos})
\bea
\sum_{n\in\bz}\,\frac{\sin n\,\theta}{n+\nu} &=& \left\{\begin{array}{lcc}
\pi\,\frac{\sin((2m+1)\pi-\theta)\nu}{\sin\pi\nu} &\;\;\;\;\;,& m2\pi< \theta<(m+1)2\pi\cr
0 &\;\;\;\;\;,& \theta = m2\pi\end{array}\right.\cr
\sum_{n\in\bz}\,\frac{\cos n\,\theta}{n+\nu} &=&
\pi\,\frac{\cos((2m+1)\pi-\theta)\nu}{\sin\pi\nu}\;\;\;\;,\;\;\;\;m2\pi\leq \theta\leq(m+1)2\pi
\;\;\;\;.
\eea
}
\be
[Z^{(a)}(\tau,\sigma) ; Z^{(b)}{}^\dagger (\tau,\sigma')] =2\;\delta^{ab}\;\left\{
\begin{array}{lcc} \Theta^{(a)}(\sigma)= \pi\,\alpha'\;\,\frac{1+ b_0^{(a)}\,b_\pi^{(a)}}
{b_\pi^{(a)} -b_0^{(a)}}\,\frac{1-b_\sigma^{(a)}{}^2}{1+b_\sigma^{(a)}{}^2}&,& \
\sigma=\sigma'=0,\pi \cr 0 &,&{\it otherwise}\end{array}\right.
\ee
signal, in the context of the paper, the non-commutative character of the effective
field theory under consideration \cite{seiwit}.
More precisely, the obvious relations $\;b_0^{(a)}=\pm \, b_\pi^{(a)}\equiv b^{(a)}$
yields the well defined non-commutative parameters
\be
\Theta^{(a)} = \frac{\pi\,\alpha'}{2}\; \frac{(1- b^{(a)}{}^2)^2}{ b^{(a)}\;(1+b^{(a)}{}^2)}
\;\;\;\;.\label{ncp}
\ee
Therefore, we are led to conjecture that the effective field theory is a five dimensional
non-commutative $U(1)\times U(1)$ gauge theory, coupled to one matter hypermultiplet charged
under one of the $U(1)'s$, with non zero non-commutative parameters in the spatial planes
$(12)$ and $(34)$, defined in (\ref{ncp}) for $a=1$ and $a=2$ respectively.
However, when SUSY holds, we have $b^{(1)}b^{(2)}= \pm 1$ and, then,
$\Theta^{(1)} = \pm\Theta^{(2)}$ follows.
For the special case $|b^{(a)}|=1$ (or what is the same, $|\nu^{(a)}| = \frac{1}{2}$)
there is a naive enhancement of the space-time symmetry from $SO(2)\times SO(2)$ to
$U(2)$, as can be checked from the spectrum and, moreover, the non-commutative parameters
are null.
Does this mean that the effective gauge field theory becomes a commutative one?
In any case, the kind of SUSY gauge theory that emerges merits further investigation, maybe
along the lines of the boundary string field theory approach \cite{bsft}.

As a last comment, it is possible that more general configurations of gauge fields give rise
to tachyon-free brane-antibrane systems preserving some supersymmetries.
In particular, we could consider  gauge fields in the brane and antibrane which were not
necessary parallel as it was recently pursued in reference \cite{choetal} for $p=2$.
A general analysis for $p\geq 2$, even though it becomes harder, is also a subject for
future work.

\appendix\section{Conventions}
\cleqn

We review shortly free field theories on the strip $\Sigma = \{\sigma^1\equiv
\sigma\in[0,\pi], \sigma^0\equiv \tau\in\br$\} with superconformal symmetry.
We adopt the following conventions.
The Minkowskian metric in two dimensions is
\be
\eta= \eta_{\mu\nu}\,d\sigma^\mu\,d\sigma^\nu = -d^2\tau + d^2\sigma = -
d\sigma^+\,d\sigma^- \;\;\;,\;\;\; \sigma^{\pm}= \tau\pm\sigma\,\;\; ,\;\;
2\,\partial_\pm = \partial_\tau\pm\partial_\sigma\;\;,
\ee
while the two-dimensional gamma matrices in a Majorana-Weyl basis are
\be
\gamma^0= -i\sigma_2 \;\;,\;\; \gamma^1=\sigma_1\;\;,\;\; \gamma^2=i\gamma^0 =\sigma_2
\;\;,\;\;\gamma_3= i \gamma^1\gamma^2 = -\sigma_3\;\;\;,\;\;\;
\{\gamma_\mu;\gamma_\nu\} = 2\,\eta_{\mu\nu}\;\;,
\ee
being $\;\frac{1}{2}\gamma_{01}=\frac{1}{2}\sigma_3\;$ the standard Lorentz generator in this
spinor representation.
We write a generic (Dirac) spinor as
\be
\psi\equiv \left(\begin{array}{c}\psi^+\cr\psi^-\end{array}\right) =
A^{-1}\;\left(\begin{array}{c}\psi_+\cr\psi_-\end{array}\right) =
\left(\begin{array}{c}\psi_-\cr\psi_+\end{array}\right)\;\;\;\;,
\ee
where we use $A=\sigma_1,A\,\gamma^\mu \,A^{-1} = +\gamma^\mu{}^t$, to low (and $A^{-1}$
to raise) spinor indices.
As  usual $\bar\psi\equiv \psi^\dagger C= i(\psi_+^*\;-\psi_-^* )$ with
$C=\sigma_2\;,\; C\gamma^\mu C^{-1}= -{\gamma^\mu}^\dagger$ the charge conjugation
matrix, while the conjugate spinor is defined by $\psi^c \equiv D\psi^*$ where
$D^{-1}\, \gamma^\mu\, D = \gamma^\mu{}^*$.
In the representation chosen, where the $\gamma^\mu$'s are real, we can take $D=1$ and then a
Majorana spinor which, by definition, verifies the reality constraint $\psi^c =\psi$, is just
a real one $\psi_\pm^* = \psi_\pm$.
We will be using the scales $T_s{}^{-1} = 2\,\pi\,\alpha' =
\pi\,l^2$ and the notation $\;\bz_\nu \equiv \bz + \nu = \{m+\nu\,,\, m\in\bz
\}\;\;,\;\;\bz'_\nu = \bz_\nu - \{ 0\}\,$ with $\nu\in[0,1)$.

To begin with, we consider the action $S[X,\psi] = S^X_0[X] + S^\psi[\psi]$ with $X ,
\psi\,$ real and
\bea
S^X_0[X] &=&-\frac{T_s}{2}\; \int_\Sigma\; d^2\sigma\;
\eta^{\mu\nu}\;\partial_\mu X \;\partial_\nu X = 2\,
T_s\,\int_\Sigma\;d^2\sigma\;\partial_+ X\;\partial_- X\cr
S^\psi[\psi] &=&-\frac{T_s}{2}\;\int_\Sigma\;d^2\sigma\; \bar\psi\;\gamma^\mu\,
\partial_\mu\psi= i\;T_s\;\int_\Sigma\;d^2\sigma\; (\psi_+\;\partial_-\psi_+ +
\psi_-\;\partial_+\psi_-)\;\;.\label{accion}
\eea
Formally, the action is invariant under conformal SUSY
transformations with real Grasmman parameters $\epsilon^{\pm} = \epsilon^{\pm}
(\sigma^{\pm})$
\bea
\delta_\epsilon X &=& \bar\epsilon\;\psi = -i\,( \epsilon^+\psi_+ - \epsilon^-\psi_-)\cr
\delta_\epsilon\psi &=&\partial_\mu X\;\gamma^\mu\; \epsilon =
\left(\begin{array}{r}-2\,\partial_-X\;\epsilon^-\cr
2\,\partial_+X\;\epsilon^+\end{array}\right)\;\;\;\;.\label{SUSY}
\eea
The equations of motion derived from (\ref{accion}) are solved by
\be
\partial_\pm X = F_\pm (\sigma^\pm)\;\;\;\;,\;\;\;\; \psi_\pm = f_\pm
(\sigma^\pm)\;\;\;\;,
\ee
for arbitrary functions $\;F_\pm, f_\pm\;$, and the cancellation of the boundary terms
\vfill\eject
\bea
\delta S_0^X|_{b.t.} &=& -T_s\;\int\;d\tau\; \delta X\;
\partial_\sigma X|_{\sigma=0}^{\sigma=\pi}\cr
\delta S^\psi|_{b.t.} &=& i\,\frac{T_s}{2}\int\;d\tau\;(\delta\psi_+\;\psi_+ -
\delta\psi_-\;\psi_-)|_{\sigma=0}^{\sigma=\pi}\label{bt1}
\eea
in the variation of the action imposes the boundary conditions (b.c.) on the fields as well
as the form of the surviving supersymmetry.
It is clear that fermions must in general satisfy the following b.c.
\be
\psi_-|_{\sigma=0} = \delta_0\;\psi_+|_{\sigma=0}\;\;\;,\;\;\; \psi_-|_{\sigma=\pi} =
\delta_\pi\;\psi_+|_{\sigma=\pi}\;\;\;\;.\label{fermionbc}
\ee
For Majoranas the phases are simply signs, $|\delta_0|=|\delta_\pi|=1$; the general solution
is
\bea
\psi_\pm(\sigma_\pm) = \left.\begin{array}{l}1\cr \delta_0\end{array} \right\}\;
\frac{l}{\sqrt{2}}\; \sum_{r\in \bz_\nu}\;b_r\; e^{-ir\sigma^\pm}\;\;\;\;,\;\;\;
\nu=\left\{\begin{array}{l}0\;\;\;,\;\;\; \delta_\pi\,\delta_0 = +1\cr
\frac{1}{2}\;\;\;,\;\;\; \delta_\pi\,\delta_0 = -1\end{array} \right.
\eea
from which we obtain
\be
b_r = \frac{1}{\sqrt{2}\pi l}\;\int_0^\pi\;d\sigma\;\left(e^{ir\sigma^+}\;\psi_+(\sigma_+)
+ \delta_0\;e^{ir\sigma^-}\;\psi_-(\sigma_-)\right)\;\;\;\;.
\ee
By definition in Ramond (R) sectors bosons and fermions have equal modding while, in
Neveu-Schwarz (NS) sectors, the modding differs by a half-integer.
So, they will differ according to the four possible choices for the
boson b.c. to be considered.

\subsection{NN boundary conditions}

Neumann b.c. are considered at both ends,
\be
\partial_\sigma X |_{\sigma = 0}=\partial_\sigma X |_{\sigma = \pi}= 0\;\;\;\;.
\ee
In this case, the general solution for the boson field is
\bea
X(\tau,\sigma) &=& x
+ l^2\, p\; \tau + i\,l\;\sum_{m\in\bz'}\; \frac{\alpha_m}{m}\; e^{-im\tau}\;\cos
m\,\sigma\cr
\partial_\pm X(\tau,\sigma) &=& \frac{l}{2}\;\sum_{m\in\bz}\;\alpha_m\; e^{-im\sigma^\pm}
\;\;\;\;,\;\;\;\;\alpha_0\equiv l\, p\label{nn}
\eea
while, for the fermions, compatibility with the superconformal symmetry (\ref{SUSY})
yields the phases $\delta_0 = 1$ and $\delta_\pi =\pm 1$ in the R/NS sectors.

\subsection{DD boundary conditions}

Dirichlet b.c. are considered at both ends,
\be
\partial_\tau X |_{\sigma = 0}=\partial_\tau X |_{\sigma = \pi}= 0\;\;\;\;.
\ee
In this case, the general solution for the boson field is ($\Delta x\equiv x_\pi -x_0$)
\bea
X(\tau,\sigma) &=& x_0 + \frac{\Delta x}{\pi}\;\sigma +
l\;\sum_{m\in\bz'}\;\frac{\alpha_m}{m}\; e^{-im\tau}\;\sin m\,\sigma\cr
\partial_\pm X(\tau,\sigma) &=& \pm\,\frac{l}{2}\;\sum_{m\in\bz}\;\alpha_m\;
e^{-im\sigma^\pm}\;\;\;\;,\;\;\;\; \alpha_0\equiv \frac{\Delta x}{\pi\,l}\equiv l\,p
\label{dd}
\eea
while, for the fermions, compatibility with SUSY transformations yields
the phases $\delta_0=-1$ and $\delta_\pi = \pm\, 1$ in the NS/R sectors.

\subsection{ND boundary conditions}

In this case, Neumman b.c. are taken at one end and Diritchlet b.c. at the other one,
\be
\partial_\sigma X |_{\sigma = 0}=\partial_\tau X |_{\sigma = \pi}= 0\;\;\;\;.
\ee
The general solution for the boson field is
\be
X(\tau,\sigma) = x_\pi + i\,l\;\sum_{r\in\bz_{\frac{1}{2}}}\;\frac{\alpha_r}{r}\;
e^{-ir\tau}\; \cos r\,\sigma\;\;\;,\;\;\;\partial_\pm X(\tau,\sigma) =
\frac{l}{2}\;\sum_{r\in\bz_\frac{1}{2}}\;\alpha_r\; e^{-ir\sigma^\pm}
\ee
while, for the fermions, compatibility with SUSY transformations yields
the phases $\delta_0=+1$ and $\delta_\pi = \pm\, 1$ in the NS/R sectors.

\subsection{DN boundary conditions}

This case is the same as the ND one, with the ends interchanged,
\be
\partial_\tau X |_{\sigma = 0}= \partial_\sigma X |_{\sigma = \pi}= 0\;\;\;\;.
\ee
The general solution for the boson field is
\be
X(\tau,\sigma) = x_0 + l\;\sum_{r\in\bz_{\frac{1}{2}}}\;\frac{\alpha_r}{r}\;
e^{-ir\tau}\; \sin r\,\sigma\;\;\;,\;\;\;\partial_\pm X(\tau,\sigma) = \pm\,
\frac{l}{2}\;\sum_{r\in\bz_\frac{1}{2}}\;\alpha_r\; e^{-ir\sigma^\pm}
\ee
while, for the fermions, compatibility with SUSY transformations yields
the phases $\delta_0=-1$ and $\delta_\pi = \pm\, 1$ in the R/NS sectors.
\bigskip

In all the four cases, the SUSY parameters are characterized by the conditions
\be
\epsilon^+(\tau)  = - \epsilon^-(\tau)\equiv\epsilon(\tau) \;\;\;\;,\;\;\;\;
\epsilon(\tau +2\pi)=\pm\epsilon(\tau)\;\;\; R/NS\; sector\;\;\;\;.
\ee
In particular the surviving global SUSY charge (that certainly lives in the R sector)
is the combination $\;Q_+ + Q_-\sim G_0$ (see (\ref{scurr})).

\subsection{Quantization and $N=1$ superconformal algebra}

From the action (\ref{accion}) we read the canonical (anti) commutation relations
\bea
[X(\tau,\sigma); \partial_\tau X(\tau,\sigma')] &=& \frac{i}{T_s}\;\delta(\sigma -
\sigma') \;\;\;\longleftrightarrow\;\;\;[\alpha_m ; \alpha_n] = m\,\delta_{m+n,0}\cr
\{\psi_\pm(\tau,\sigma); \psi_\pm(\tau,\sigma')\} &=& \frac{1}{T_s}\;\delta(\sigma -
\sigma') \;\;\;\longleftrightarrow\;\;\; \{b_r ; b_s\} = \delta_{r+s,0}\;\;\;\;. \label{ccr}
\eea
Reality conditions translate into $\alpha_{-m}{}^\dagger =
\alpha_m\;,\;b_{-r}{}^\dagger = b_r$.
Quantum-mechanically, a normal ordering prescription is needed to define quantum operators;
this is done from the usual representation of the canonical conmutation relations (\ref{ccr})
\footnote{
For periodic (P) fermions, if we identify $|0>\equiv |->$.
Then, $\;b_0 |\pm> = \frac{1}{\sqrt 2}\,|\mp>\;$.
}
\be
|0>\;\;:\;\;\;  \alpha_m |0> =  b_m |0> = 0 \;,\; m>0\label{vd}
\ee
by putting at right (left) the destruction (creation) operators.
The (traceless) energy-momentum tensor is defined by
$T_{\mu\nu}\equiv-\frac{2}{T_s}\,\frac{\delta S}{\delta g^{\mu\nu}}|_{g=\eta}$;
its non zero components for bosons and fermions are, respectively,
\bea
T_{\pm\pm}^X(\sigma^\pm) &=& \partial_\pm
X \;\partial_\pm X \equiv\alpha'\;\sum_{m\in\bz}\;L_m^X\; e^{-im\sigma^\pm}\cr L_m^X
&=& \frac{1}{2}\sum_{n}\; :\alpha_{m-n}\;\alpha_n: + \Delta_0^X\,\delta_{m,0}
\;\;\;,\;\;\; [L_m^X;\alpha_n] = -n\,\alpha_{m+n}\cr
T_{\pm\pm}^\psi(\sigma^\pm)
&=& \frac{i}{2}\;\psi_\pm\;\partial_\pm\psi_\pm \equiv\alpha'\;\sum_{m\in\bz}\;L_m^\psi
\; e^{-im\sigma^\pm}\cr L_m^\psi &=& \frac{1}{2}\sum_{r}\; (r-\frac{m}{2})\; :b_{m-r}
\;b_r : + \Delta_0^\psi\,\delta_{m,0} \;\;\;,\;\;\;[L_m^\psi; b_r] = -(r +
\frac{m}{2})\, b_{m+r}\;\;.\cr & &\label{vir}
\eea
The Virasoro generators $\{L_m^X\}$ and $\{L_m^\psi\}$ obey the algebra in (\ref{n1alg})
with $c^X=1$ and $c^\psi=\frac{1}{2}$ respectively, provided that the conformal
dimensions of the vacuum states are
\footnote{
Alternatively they can be computed by using the
Hurwitz zeta-function $\xi(s,x) \equiv \sum_{k=0}^\infty\; (k+x)^{-s}\;$ (particularly
useful cases are $\;\xi(-1,x)= -\frac{1}{12}\,(1 + 6\,x^2 - 6\,x)$ and $\;\xi(-2,x)=
-\frac{1}{6}\, x\,(1-x)\,(1-2x)$) to regulate the infinite sums and then adding the
Casimir energy $\frac{c}{24}$, see \cite{polcho1}.
}
\be
\Delta_0^X = \left\{\begin{array}{rcc} -\frac{1}{24} + \frac{1}{24} = &0& \mbox{NN ,
DD}\cr +\frac{1}{48} + \frac{1}{24}= &\frac{1}{16}& \mbox{ND ,
DN}\end{array}\right.\;\;\;\;,\;\;\; \Delta_0^\psi = \left\{\begin{array}{rcc}
-\frac{1}{48} + \frac{1}{48}= &0& \mbox{AP}\cr +\frac{1}{24} + \frac{1}{48} =
&\frac{1}{16}& \mbox{P}\end{array}\right.\;\;\;.\label{zmc}
\ee
Similarly, the fermionic supercurrent defined by
$G_\mu \equiv -\frac{1}{2T_s}\,\frac{\delta S}{\delta \chi^\mu}|_{\chi = 0 }\;$,
where $\chi^\mu $ is the gravitino field, has the components
\bea
G_\pm (\sigma^\pm) &=& \psi_\pm\;\partial_\pm X \equiv \frac{\alpha'}{\sqrt{2}}\;
\sum_{r\in \bz_\delta}\; G_r\; e^{-ir\sigma^\pm}\;\;\;,\;\;\;
\delta=\left\{\begin{array}{l}0\;\;,\;\;R\cr\frac{1}{2}\;\;,\;\;NS\end{array}\right.\cr
G_r &=& \sum_m \; \alpha_m\; b_{r-m}\;\;\;,\;\;\; \{G_r;b_s\}=\alpha_{r+s}\;\;\;,\;\;\;
[G_r;\alpha_m]=-m\,b_{r+m}\;\;.\label{scurr}
\eea

In the NN or DD (with $x_0=x_\pi$) cases, the NS vacuum defined in (\ref{vd}) is the unique
$Osp(1,2)$ (SUSY extension of $SL(2,\br)$) invariant one,
\be
L_m\, |0>_{NS} =  G_r \, |0>_{NS} = 0  \;\;\;,\;\;\; m\geq -1 \;\;,\;\;
r\geq-\frac{1}{2}\;\;\;\;.
\ee
This is not so with ND/DN b.c. (or DD with $\Delta x\neq 0$)
because $L_{-1}$ does not annihilate it; for example, $L_{-1}^X |0>_{NS} = \frac{1}{2}
\alpha_{-\frac{1}{2}}{}^2  |0>_{NS}(\frac{\Delta x}{\pi\, l}\, \alpha_{-1}|0>_{NS}$ ).

As we saw, the combined system in each case present two sectors, the NS sector with
opposite modding and the R sector with equal modding.
The standard form of the $N=1$ superconformal algebra
\bea
[L_m; L_n] &=& (m-n)\;  L_{m+n} + \frac{c}{12}\; m\;(m^2-1)\; \delta_{m+n,0}\cr
[L_m;  G_r] &=& (\frac{m}{2}-r)\; G_{m+r}\cr \{G_r; G_s\} &=&
2\, L_{r+s} + \frac{c}{12}\; (4\, r^2 - 1)\; \delta_{r+s,0}\;\;\;\;,\label{n1alg}
\eea
where $c=\frac{3}{2}$ is the central charge of the system, follows, with
\be
\Delta^{(R)}_0= \frac{1}{16}\;\;\;,\;\;\; \Delta^{(NS)}_0 = \left\{\begin{array}{l}
0\;\;\; \mbox{NN, DD}\cr \frac{1}{8}\;\;\;\mbox{ND, DN}\end{array}\right.\;\;\;\;.\label{ve}
\ee

\section{A system with mixed boundary conditions}
\cleqn

Let us take a complex scalar field $Z = X^1 + i\, X^2$, together with a Dirac fermion
$\Psi=\psi^1 + i\, \psi^2$ with $X^i, \psi^i$ real, and consider the action $S = S_0^Z +
S^\Psi + S_b^Z$, where
\bea
S^Z_0 &=&-\frac{T_s}{2}\; \int_\Sigma\; d^2\sigma\;\eta^{\mu\nu}\;
\partial_\mu Z^* \;\partial_\nu Z = S_0^{X^1} + S_0^{X^2}\cr
S^\Psi &=& -\frac{T_s}{4}\;\int_\Sigma\;d^2\sigma\;
(\bar\Psi\;\gamma^\mu\,\partial_\mu\Psi + h.h.)= S^{\psi^1}+S^{\psi^2}\cr
S_b^Z &=&\frac{T_s}{2}\;\int\;d\tau\; \epsilon_{ij}\; \left( b_\pi\;
X^i(\tau,\pi)\partial_\tau X^j(\tau,\pi) - b_0\; X^i(\tau,0)\partial_\tau
X^j(\tau,0)\right)\;\;.\label{accionmixed}
\eea
The boundary term $S_b$ can be interpreted as the coupling, with unit charge, of the
ends of the string to a gauge field $A$ of strength $F \equiv dA = T_s\,\tilde
b(\sigma)\; dX^1\wedge dX^2\;$, where $\;\tilde b(0) = b_0 \;,\;\tilde b (\pi) = b_\pi\;$
and zero otherwise.
\bigskip

\noindent{\underline{ The bosonic sector}}

The boundary conditions that follow from (\ref{accionmixed}) are
\be
(\partial_\sigma + i\, b_0\,\partial_\tau )Z(\tau,\sigma)|_{\sigma=0} =
(\partial_\sigma + i\, b_\pi \,\partial_\tau )Z(\tau,\sigma)|_{\sigma=\pi} = 0\;\;\;\;.
\label{bcm}
\ee
It will be convenient to introduce the following notation
\be
1+ i\,b_0 \equiv \sqrt{1+b_0{}^2}\; e^{i\varphi_0} \;\;\;,\;\;\; \varphi_0\equiv
\frac{\pi}{2}\,\nu_0\;\;\;;\;\;\; 1+ i\,b_\pi\equiv \sqrt{1+b_\pi{}^2}\;
e^{i\varphi_\pi} \;\;\;,\;\;\; \varphi_\pi\equiv \frac{\pi}{2}\,\nu_\pi\;,
\ee
where $\nu_0\, (\nu_\pi) $ varies from $-1$ to $+1$ as $b_0 \,(b_\pi)$ goes from
$-\infty$ to $+\infty$.
We also introduce
\be
\nu_\pm \equiv \frac{1}{2}\,(\nu_\pi\pm \nu_0)\;\;\;\;,\;\;\;\;\; \bar\nu =
\left\{\begin{array}{lcr}\nu_-\;\;\;&,&\;\;\; 0\leq\nu_-<1\cr\nu_- +1\;\;\;&,&\;\;\;
-1<\nu_-<0\end{array}\right.\;\;\;\;.\label{barnu}
\ee

The general solution with the boundary conditions (\ref{bcm}) can be written
\bea
Z(\tau,\sigma) &=& z +\sqrt{2}\,
l\, A_0\; \phi_0 (\tau,\sigma) + i\,\sqrt{2}\,l\;\sum_{r\in\bz'_{\bar\nu}} \;
\frac{A_r}{r}\;\phi_r(\tau,\sigma)\cr
\partial_\pm Z(\tau,\sigma) &=& \frac{l}{\sqrt{2}}\; \sum_{r\in\bz_{\bar\nu}}\;
A_r\; e^{-ir\sigma^\pm \mp i\varphi_0}\;\;\;\;,\label{sngralmixed}
\eea
where we have introduced the functions
\bea
\phi_0(\tau,\sigma) &=& \cos \varphi_0 \;\tau - i\, \sin\varphi_0
\,\sigma\cr \phi_r(\tau,\sigma) &=& e^{-ir\tau}\; \cos(r\,\sigma +
\varphi_0)\;\;\;,\;\;\; r\in\bz'_{\bar\nu}
\eea
and it is understood that $A_0\equiv 0$ unless $\bar\nu=0$.

Let us introduce, at fixed time, the pairing
\be
(\phi_1;\phi_2) \equiv \frac{1}{2i}\,\int_0^\pi\;d\sigma\; (\phi_1^*\;
\partial_\tau\phi_2 -\partial_\tau\phi_1^*\;\phi_2) +
\frac{b_\pi}{2}\;\phi_1^*\;\phi_2|_{\sigma=\pi} -
\frac{b_0}{2}\phi_1^*\;\phi_2|_{\sigma=0}\;\;\;.\label{pairing}
\ee
It is easy to prove that, for $\phi_1,\phi_2$ of the type (\ref{sngralmixed}), it does
not depend on $\tau$.
In particular, the following orthogonality relations hold
\bea
(\phi_r;\phi_s) &=& \delta_{r,s}\; \left\{\begin{array}{ll}-\frac{\pi}{2}\; r &,\;\;\;
r\neq 0\cr\frac{\pi^2}{2}\;b_0\; &,\;\;\; r=0 \end{array}\right.\cr
(1;\phi_r)&=&\frac{\pi}{2\,i}\;\sec\varphi_0\;\delta_{r,0}\cr
(1; 1)&=&\frac{\sin\pi\nu_-}{\cos\pi\nu_++\cos\pi\nu_-} = \frac{1}{2}\,(b_\pi - b_0)
\;\;\;\;.
\eea
From them, it follows that
\bea
(1;Z) &=& (1;1)\;z - \sqrt{2}\,l\;(\phi_0;1)\; A_0 \cr
(\phi_0;Z) &=& (\phi_0;1)\;z +\sqrt{2}\,l\;(\phi_0;\phi_0)\; A_0 \cr
(\phi_r;Z) &=&\frac{\pi\,l}{\sqrt{2}\,i}\; A_r \;\;\;,\;\;\; r\in \bz'_{\bar\nu}\;\;\;\;.
\label{coef}
\eea
Then, the bulk commutation relations implied by (\ref{accionmixed})
\be
[Z(\tau,\sigma); \partial_\tau Z(\tau,\sigma')] = i\,\frac{2}{T_s}\;\delta(\sigma -
\sigma')
\ee
yield, from (\ref{coef}), the non trivial commutation relations
\footnote{
An useful relation is the following one; if $f_i  = (\phi_i ; Z)\;,\;i=1,2\;$, then
$[f_1;f_2{}^\dagger] = -\frac{1}{T_s}\;(\phi_1;\phi_2)$.
}
\vfill\eject
\bea
[z;z^\dagger] &=&
2\,\theta \;\;\; \longleftrightarrow\;\;\;[x^i;x^j] =
i\,\theta\;\epsilon^{ij}\;\;\;,\;\;\; z \equiv x^1 + i\,x^2\cr [A_r; A_s{}^\dagger] &=&
r\,\delta_{r,s}\;\;\;,\;\;\; [z;A_0{}^\dagger] = i\,\sqrt{2}\,l\, \cos \varphi_0\;\;\;\;,
\eea
where the non-commutative parameter $\theta$ is given by
\be
\theta = \left\{ \begin{array}{ll}-(2\,T_s\,(1;1))^{-1}= \frac{1}{T_s}\;\frac{1}{b_0
-b_\pi} \;\;\;&,\;\;\;\bar\nu\neq 0\cr \frac{\sin\pi\nu_0}{2\,T_s} =
\frac{1}{T_s}\;\frac{b_0}{1+ b_0{}^2}\;\;\;&,\;\;\; \bar\nu=0\end{array}\right.\;\;\;\;.
\ee

\bigskip

\noindent{\underline{ The fermionic sector}}

From the action for the Dirac spinor in (\ref{accionmixed}) with the boundary
conditions as in (\ref{fermionbc}), but now with $\delta_0\equiv e^{i\phi_0}$ and
$\delta_\pi\equiv e^{i\phi_\pi}$ arbitrary phases, the general solution is given by
\be
\Psi_\pm(\sigma_\pm) = l\, \sum_{r\in \bz_{\nu}}\;B_r\; e^{-ir\sigma^\pm \mp
i\frac{\phi_0}{2}} \;\;\;, \;\;\; \delta_\pi\,\delta_0^* = e^{i(\phi_\pi-\phi_0)}\equiv
e^{i2\pi\nu}\;\;\;\;.\label{fermixed}
\ee
However, consistency with $N=1$ superconformal symmetry transformations
\footnote{
Equivalently, it is possible to derive the b.c. from a fermionic boundary term added to
the action (\ref{accionmixed}), see for example \cite{bp}.
}
\bea
\delta_\epsilon\,Z &=& \bar\epsilon\;\Psi = -i\,( \epsilon^+\Psi_+ - \epsilon^-\Psi_- )\cr
\delta_\epsilon\Psi &=&\partial_\mu Z\;\gamma^\mu\; \epsilon =
\left(\begin{array}{r}-2\,\partial_-Z\;\epsilon^-\cr2\,\partial_+Z\;\epsilon^+\end{array}
\right)\;\;\;\;,
\eea
where the parameter $\epsilon$ is Majorana, fixes the phases to be
\be
\delta_0 = e^{i\pi\nu_0} \;\;\;,\;\;\;\delta_\pi = \epsilon\;e^{i\pi\nu_\pi}\;\;\;,
\;\;\; \delta_\pi\;\delta_0^* = \epsilon\;e^{i2\pi\nu_-}\;\;\;\;,\label{fasesmix}
\ee
with $\epsilon$ an arbitrary sign, as well as the form of the SUSY parameters
$\epsilon^\pm(\sigma^\pm)$
\be
\epsilon^+(\tau) = -\epsilon^-(\tau)\equiv \epsilon(\tau)\;\;\;,\;\;\; \epsilon(\tau
+2\pi) =  \epsilon\;\epsilon(\tau)\;\;\;\;.\label{epsilon}
\ee
It is easy to see that we can identify the sector with $ \epsilon = -1$ as the NS sector
and that with $ \epsilon=+1$ with the R one, just by realizing, from (\ref{fermixed}) and
(\ref{fasesmix}), that, in the last case, the fermionic modding coincides with the bosonic
one ($\nu = \bar\nu$) while, in the first one, the modding differs by a half-integer
($\nu = \bar\nu \pm \frac{1}{2}\;$ if
$\;0\leq \bar\nu < \frac{1}{2}\,/\,\frac{1}{2}\leq\bar\nu < 1 $); furthermore, from
(\ref{epsilon}) the right periodicity for the SUSY parameter follows.

Now, by using
\be
B_r = \frac{1}{2\,\pi\,l}\; \int_0^\pi \;d\sigma\; \left( e^{i(r\sigma^+ +
\varphi_0)}\;\Psi_+(\sigma^+) +
 e^{i(r\sigma^- - \varphi_0)}\;\Psi_-(\sigma^-)\right)\;\;\;\;,
\ee
the canonical anti-commutation relations read
\be
\{\Psi(\tau,\sigma); \Psi^\dagger(\tau,\sigma')\} = \frac{2}{T_s}\;\delta(\sigma -
\sigma')\;\;\;\longleftrightarrow \;\;\; \{B_r ;B_s^\dagger\} = \delta_{r,s}\;\;\;\;.
\ee
A reference state, w.r.t. the normal ordering will be understood, is defined by the
conditions
\bea
|0>_{r_0} \;:\; A_r\,|0>_{r_0}&=& 0 \;\;,\;\; r\geq 0\;\;\;,\;\;
A_r{}^\dagger\,|0>_{r_0}= 0 \;\;,\;\; r\leq 0\cr |0>_{r_0} \;:\; B_r\,|0>_{r_0}&=& 0
\;\;,\;\; r\geq r_{0} \;\;,\;\; B_r{}^\dagger\,|0>_{r_0}= 0 \;\;,\;\; r< r_{0}
\label{msvac}
\eea
for some $r_0\,\in\bz_\nu$.
\bigskip

\subsection{Conserved currents and $N=2$ superconformal algebra}

Two conserved currents associated with this system can be defined. The first one is just
the momentum associated to the translations $Z\rightarrow Z + c\;,\;c\in\com\;$, and it
is defined by
\footnote{
The reader notes that for
$|\nu_0|=1\;(|b_0|\rightarrow\infty)$ {\it or} $|\nu_\pi|=1\; (|b_\pi|\rightarrow\infty)$
it does not exist, in agreement with the fact that (\ref{bcm}) becomes Diritchlet boundary
conditions in at least one end.
}
\bea
P^\mu_i(\tau,\sigma) &\equiv&-T_s\;(\partial^\mu X_i(\tau,\sigma) - \epsilon_{ij}\;
\epsilon^{\mu\nu}\;\partial_\nu(\tilde b(\sigma)\;  X^j(\tau,\sigma)
)\;\;\;,\;\;\;\epsilon_{\tau\sigma} = \epsilon^{\sigma\tau}\equiv +1\cr p_z
&\equiv& p_1 + i\, p_2 = \int_0^\pi\;d\sigma\; P^\tau_z(\tau,\sigma) =
\left\{\begin{array}{ll}\frac{1}{i\, \theta}\; z \;\;\; &,\;\;\;\bar\nu\neq 0\cr
\frac{\sqrt{2}}{l\,\cos\varphi_0}\;A_0  \;\;\; &,\;\;\;\bar\nu=0\end{array}\right.\;\;\;\;.
\eea
They are, as they should, the canonical conjugate variables to the center of mass
coordinates,
\be
[x^i;x^j] = i\,\theta\,\epsilon^{ij} \;\;\;,\;\;\;  [x^i;p_j]= i\,\delta^i_j
\;\;\;,\;\;\; [p_i;p_j]= 0\;\;\;\;.
\ee
We can get a representation of this zero mode algebra
(for $\nu_- =0$) in momentum space by defining the action of the operators as
\be
\hat p _i \longrightarrow p_i\;\;\;, \;\;\;\hat x^i \longrightarrow
i\,\frac{\partial}{\partial p_i} - \theta\,\epsilon^{ij}\;p_j
\ee
or, in a more standard way, we can introduce canonical variables defined by
\bea
q^1 &=& c\; x^1 - s\;p_2\;\;\;\;,\;\;\;\;  k_1 = c\; x^2 + s\; p_1\cr q^2 &=& s\; x^1 + c\;
p_2\;\;\;\;,\;\;\;\;  k_2 = -s\; x^2 + c\; p_1\;\;\;\;,
\eea
where $\;c\equiv\frac{1}{\sqrt{1+\theta_2{}^2}}\;,\; s \equiv
\frac{\theta_2}{\sqrt{1+\theta_2{}^2}}\;$, $\theta_{1/2} \equiv \sqrt{1 +
\frac{\theta^2}{4}} \pm \frac{\theta}{2}>0\;$.
They verify
\bea
[q^i;q^j] = [k_i; k_j]=0 \;\;\;&,&\;\;\;  [q^i;k_j]= i\,\theta_i\;\delta^i_j\cr dx^1\wedge dx^2
\wedge dp_1\wedge dp_2 &=& dq^1\wedge dq^2 \wedge dk_1\wedge dk_2\;\;\;\;.
\eea
In view of the relation $\;\theta_1\,\theta_2 =1\;$, the phase space volume remains invariant,
an important fact in computing the zero mode contribution to the partition function.
On the other hand, when $\bar\nu\neq 0$ we have just the $x$-variables, introducing a factor
of $(2\,\pi\,\theta)^{-1}$ in such computation.

The second conserved current $J_\mu \equiv J_\mu^Z + J_\mu^\Psi$ leads to the conserved
charge $Q = Q^Z + Q^\Psi$ that generates the $U(1)$ rotations $Z\rightarrow
e^{-i\epsilon}\;Z\;,\;\Psi\rightarrow e^{-i\epsilon}\;\Psi\;$, and it is defined by
\bea
J_\mu^Z (\tau,\sigma) &\equiv& T_s\;Im\left(Z^*(\tau,\sigma)\;\partial_\mu
Z(\tau,\sigma)\right) + \frac{T_s}{2}\;\epsilon_{\mu\nu}\;\partial^\nu(\tilde
b(\sigma)\;Z^*(\tau,\sigma)\;Z(\tau,\sigma))\cr
&=& X^2 P_1^\mu - X^1 P_2^\mu\cr Q^Z
&\equiv& \int_0^\pi d\sigma\; J^Z{}^\tau(\tau,\sigma) =
\sum_{r\in\bz'_{\nu_-}}\frac{:A_r^\dagger\,A_r:}{r}\; + \left\{\begin{array}{ll}
\frac{\vec x^2}{2\,\theta} &,\; \nu_-\neq 0\cr -\frac{\theta}{2}\;\vec p^2 - x^1\, p_2
+ x^2\, p_1  &,\; \nu_-=0\end{array}\right.\cr
& &\cr J_\mu^\Psi (\tau,\sigma) &\equiv&
i\,\frac{T_s}{2}\;\bar\Psi\,\gamma_\mu\,\Psi\cr Q^\Psi &\equiv& \int_0^\pi\;d\sigma\;
J^\Psi{}^\tau(\tau,\sigma) = \frac{T_s}{2}\;\int_0^\pi\;d\sigma\;
(\Psi_+^\dagger\;\Psi_+ + \Psi_-^\dagger\;\Psi_-) = \sum_{r\in\bz_{\nu}}
:B_r^\dagger\,B_r: \;\;\;.\cr & &\label{rotcurrent}
\eea
While the bosonic part of the $U(1)$ current suggested us to introduce the pairing
(\ref{pairing}), the fermionic part can be extended to a holomorphic current,
\bea
J_\pm(\sigma_\pm ) &\equiv&
i\,\frac{T_s}{2}\;\bar\Psi\,\gamma_\pm\,\Psi =
-\frac{T_s}{2}\;\Psi_\pm^\dagger\,\Psi_\pm \equiv -\frac{1}{2\pi}\;
\sum_{m\in\bz}\;J_m\;e^{-im\sigma^\pm}\;\;\;,\;\;\; J_{-m}^\dagger = J_m\cr J_m
&=&\sum_{r\in \bz_\nu}\; :B_r^\dagger\,B_{m+r}: + q_0\,\delta_{m,0}\;\;\;,\;\;\;
[J_m;B_r]=-B_{m+r}\label{u1cur}
\eea
whose zero component is essentially the global charge defined above,
$J_0 = Q^\Psi + q_0$,
where $q_0\equiv r_0 -\frac{1}{2}$ is the $U(1)$ charge of $|0>_{r_0}$.
This current is just the $U(1)$ current of the $N=2$ superconformal algebra the system
realizes, the fermion number current.
In fact, we can introduce the {\it complex} fermionic currents
\bea G_\pm(\sigma^\pm) &\equiv& \frac{1}{2}\;
\Psi_\pm^*\;\partial_\pm Z \equiv \frac{\alpha'}{\sqrt{2}}\, \sum_{r\in\bz_\delta}\;
G_r^{(+)}\; e^{-ir\sigma^\pm} \;\;,\;\;\delta=\left\{\begin{array}{l} 0\;\;,\;R\cr
\frac{1}{2}\;\;,\;NS\end{array}\right.\cr G_r^{(+)}
&\equiv&\sum_{s\in\bz_{\bar\nu}}\;A_s\;B_{s-r}^\dagger \;\;\;,\;\;\;
[G_r^{(+)};A_s^\dagger]=
s\,B_{s-r}^\dagger\;\;\;,\;\;\;\{G_r^{(+)};B_s\}=A_{r+s}\;\;.\label{n2fermcur}
\eea
The modes of the hermitian conjugate $J_\pm(\sigma^\pm)^\dagger$ are introduced in the
same way, with the result $\;G_r^{(-)}= G_{-r}^{(+)}{}^\dagger$.
Furthermore, the modes of the bosonic and fermionic energy-momentum tensors
\bea
T_{\pm\pm}^Z &\equiv& \partial_\pm
Z^\dagger\;\partial_\pm Z \equiv \alpha'\;\sum_{m\in\bz}\;L_m^Z\; e^{-im\sigma^\pm}\cr
L_m^Z &=& \sum_{r\in\bz_{\bar\nu}}\;:A_r^\dagger\,A_{r+m}: + \Delta_0^Z\,\delta_{m,0}
\;\;\;,\;\;\;[L_m^Z;A_r]= -r\, A_{m+r} \cr T_{\pm\pm}^\Psi &\equiv&\frac{i}{4}\;
(\Psi_\pm^\dagger\,\partial_\pm\Psi_\pm +\Psi_\pm\,\partial_\pm\Psi_\pm^\dagger)\equiv
\alpha'\;\sum_{m\in\bz}\;L_m^\Psi\;e^{-im\sigma^\pm}\cr
L_m^\Psi &=& \sum_{r\in\bz_\nu}\, (r +\frac{m}{2})\, :B_r^\dagger\,B_{m+r}: +
\Delta_0^\Psi\,\delta_{m,0} \;\;,\;\; [L_m^\Psi;B_r]= -(r+\frac{m}{2})\, B_{m+r}
\label{emtms}
\eea
verify the standard Virasoro algebra in (\ref{n1alg}) with $\;c^Z = 2\;$,
$\;c^\Psi = 1,\;$ respectively, and the constants
\be
\Delta_0^Z = \frac{1}{8} - \frac{1}{2}\,(\bar\nu -\frac{1}{2})^2\;\;\;\;,\;\;\;\;
\Delta_0^\Psi = \frac{1}{2}\,(r_0 -\frac{1}{2})^2\label{vems}\;\;\;\;.
\ee
Thus, the modes of the mixed system, $L_m = L_m^Z + L_m^\Psi$ obey the Virasoro algebra
with $\;c = 3$ and a conformal dimension of the vacuum given by
\be
\Delta = \Delta^Z_0 + \Delta^\Psi_0 = \frac{1}{8}+
\frac{1}{2}\,(r_0-\bar\nu)(r_0 +\bar\nu -1)\;\;\;\;.
\ee
This energy-momentum tensor, together with the currents (\ref{u1cur}),
(\ref{n2fermcur}), generate the $N=2$ superconformal algebra which completes the
Virasoro one with
\bea
\{G_r^{(+)};G_s^{(-)} \} &=& L_{r+s} + \frac{r-s}{2}\;J_{r+s} +
\frac{c}{24}\; (4\,r^2-1)\; \delta_{r+s,0}\cr [J_m;J_n] &=& \frac{c}{3}\;
m\;\delta_{m+n,0}\cr [L_m;J_n] &=& -n\, J_{m+n}\cr [L_m;G_r^{(\pm)}] &=&
(\frac{m}{2}-r)\, G_{m+r}^{(\pm)}\cr [J_m;G_r^{(\pm)}] &=& \pm\;G_{m+r}^{(\pm)}\;\;\;\;.
\eea
It is easy to check that the current $G_\pm + G_\pm^\dagger$, with modes $G_r = G_r^{(+)}
+G_r^{(-)}$, closes in the $N=1$ superconformal algebra (\ref{n1alg}), which is an
important fact because it is this hermitian current that enters in the super Virasoro
constraints.

All this is true for any choice of $r_0\in\bz_\nu$ in (\ref{msvac}).
The choice
\be
r_0 = \bar\nu + \delta\label{r0}
\ee
defines states with minimal conformal dimension, i.e. vacuums, except in the
NS sector, when $\frac{1}{2}\leq\bar\nu < 1$, where such a state is
$\;B_{\bar\nu - \frac{1}{2}}|0>_{\bar\nu +\frac{1}{2}} =|0>_{\bar\nu - \frac{1}{2}}\;$.
In any case, due to the identifications
\be
|0>_{r_0 + n}\equiv \left\{\begin{array}{lcr}
B_{r_0 + n}\dots B_{r_0 -1}|0>_{r_0} &,& n\in\bz^-\cr
B^\dagger_{r_0 + n-1}\dots B^\dagger_{r_0}\,|0>_{r_0} &,& n\in\bz^+
\end{array}\right.\;\;\;\;,
\ee
all the states $\{|0>_{r_0 +n}\;,\; n\in\bz\}$ are in the representation defined by
$|0>_{r_0}$; so, with no loss of generality, we will take for $r_0$ the value (\ref{r0}).
It is worth to noting that any of these vacua is $Osp(1,2)$ invariant unless $\bar\nu =0$,
in which case it belongs to the NS sector.
This follows, for example, from
$\; L_{-1}^Z |0> =  A_{\bar\nu}^\dagger\, A_{\bar\nu -1}\,|0>\;$,
and $\;L_{-1}^\Psi|0>_{r_0} = (r_0-\frac{1}{2})B^\dagger_{r_0}\,B_{r_0 -1}|0>_{r_0}$.
The superconformal invariant vacuum can be written in the bosonized form of the theory;
we do not dwell in such details here \cite{polcho2}.
\vfill\eject

\section{Ghost systems}
\cleqn

Ghost systems are generically defined as theories of fields that obey commutation
relation (statistics) opposite to the usual ones assigned according to their behaviour
under Lorentz transformations (spin) by the CPT theorem, i.e., spin integer fields are
anti-commuting, while half-integer spin fields are commuting.
They naturally appear as a Fadeed-Popov representation of determinants that come from
fixing gauge symmetries.
In superstring theories, two dimensional reparameterization invariance gives rise to the
anticommuting $\lambda=2\;$ $b$-$c$ system, while the gauge fixing of the local
world-sheet SUSY gives rise to the commuting $\lambda=\frac{3}{2}\;$ $\beta$-$\gamma$ system.
Furthermore, an anticommuting $\lambda=1\;$ $\eta$-$\xi$ system is needed in the process
of bosonization of the last ones.

\subsection{Anticommuting $b$-$c$ systems}

Let us consider a pair of anticommuting, completely symmetric, and traceless tensors
$c\in \tau^{\lambda-1}_0$ and $b\in \tau^0_\lambda$ with $\lambda\in\bz$;
in the conformal gauge they have components
$( c\equiv c^{+\dots+} , \tilde c\equiv c^{-\dots-})$ and $(b\equiv b_{+\dots+}, \tilde
b\equiv b_{-\dots -})$ respectively, with action
\be
S^{bc} = i\, T_s\;\int_\Sigma\; d^2\sigma\;\left(c\;\partial_- b + \tilde c\;\partial_+
\tilde b\right)\;\;\;\;.
\ee
The equations of motion and boundary conditions to be considered
\bea
\partial_- b &=& \partial_- c = \partial_+\tilde b = \partial_+ \tilde c = 0\cr
\tilde b |_{\sigma=0,\pi} &=& e^{i\gamma_{0,\pi}}\; b |_{\sigma=0,\pi}\;\;\; ,\;\;\;
\tilde c |_{\sigma=0,\pi} = e^{-i\gamma_{0,\pi}}\; c |_{\sigma=0,\pi}\label{obcbc}
\eea
yield the $\; \epsilon_0\equiv \frac{\gamma_\pi -\gamma_0}{2\,\pi}\;$ modded expansions
\bea
c(\sigma^+) &=& l\; \sum_{p\in\bz_{\epsilon_0}}\;c_{-p}\;e^{ip\sigma^+ +
i\frac{\gamma_0}{2}} \;\;\;,\;\;\; \tilde c(\sigma^-) = l\;
\sum_{p\in\bz_{\epsilon_0}}\;c_{-p}\;e^{ip\sigma^- -
 i\frac{\gamma_0}{2}}\cr
b(\sigma^+) &=& l\; \sum_{p\in\bz_{\epsilon_0}}\;b_p\;e^{-ip\sigma^+ -
i\frac{\gamma_0}{2}}\;\;\;,\;\;\; \tilde b(\sigma^-) = l\;
\sum_{p\in\bz_{\epsilon_0}}\;b_p\;e^{-ip\sigma^- + i\frac{\gamma_0}{2}}
\eea
while, with the help of
\bea
c_{-p} &=& \frac{1}{2\pi l}\; \int_0^\pi\; d\sigma\; \left(
e^{-i\frac{\gamma_0}{2}-ip\sigma^+} \; c(\sigma^+) +
e^{i\frac{\gamma_0}{2}-ip\sigma^-} \; \tilde c(\sigma^-)\right)\cr
b_{p} &=& \frac{1}{2\pi l}\; \int_0^\pi\; d\sigma\; \left(
e^{i\frac{\gamma_0}{2}+ip\sigma^+} \; b(\sigma^+) +
e^{-i\frac{\gamma_0}{2}+ip\sigma^-} \; \tilde b(\sigma^-)\right)\;\;\;\;,\label{bcinv}
\eea
the canonical anti-commutation relations read
\be
\{b(\tau,\sigma); c(\tau,\sigma')\} = \frac{2}{T_s}\;\delta(\sigma -\sigma')\;\;\;
\longleftrightarrow \;\;\;\{ c_{-p}\,;b_q \} = \delta_{p,q}\;\;\;\;.
\ee
We remark that the case $\epsilon_0=0$ is the one relevant to string theory, because this
is the modding of the reparameterization parameters.

A ghost reference state (loosely called ``vacuum") is defined by the conditions
\be
b_p\; |0>_{p_0} = 0 \;\;\;,\;\;\; p\geq p_0  \;\;\;\;;\;\;\;\; c_{-p}\; |0>_{p_0} = 0
\;\;\;,\;\;\;p < p_0\;\;\;\;,\label{ghostvacuum}
\ee
where $p_0\in\bz_{\epsilon_0}$.
With respect to it, we define the components of the energy-momentum tensor
(we omit the right moving expressions)
\bea
T(\sigma^+) &=& \frac{1-\lambda}{2\,i}\; \partial_+b\; c:-\frac{\lambda}{2\,i}\;
b\;\partial_+c :\equiv \alpha'\;\sum_{m\in\bz}\; L^{bc}_m\; e^{-im\sigma^+}\cr
L^{bc}_m &=& \sum_{p\in\bz_{\epsilon_0}}\;\left( (\lambda - 1)\; m -p \right)\;
:b_{m+p}\;c_{-p}: + \delta_{m,0}\;\Delta^{bc}_0\cr
[L^{bc}_m; b_p] &=& \left( (\lambda-1)\;m - p \right)\;b_{m+p}\;\;\;\;,\;\;\;\;
[L^{bc}_m; c_{-p}] = \left( -\lambda\;m +p \right)\;c_{m-p}\;\;\;\;.
\eea
Moreover, the symmetry under
$c\rightarrow e^\lambda c\;,\; b\rightarrow e^{-\lambda}\, b\;$
gives rise to a conserved ghost number current
\bea
U(\sigma^+) &\equiv& - :b\; c: = l^2\,\sum_{m\in\bz}\;U_m\;e^{-im\sigma^+}\;\;\;\;;\;\;\;\;
U_m \equiv-\sum_{p\in\bz_{\epsilon_0}}\; :b_{m+p}\;c_{-p}: + u_0\;\delta_{m,0}\cr
[U_m; b_p] &=&-b_{m+p}\;\;\;\;,\;\;\;\; [U_m; c_{-p}] = +c_{m-p}\;\;\;\;.\label{gnc}
\eea
The operators $\;L^{bc}_m,\; U_m\;$, satisfy the standard algebra (with $\epsilon =+1$)
\bea
[L^{bc}_m; L^{bc}_n] &=& (m-n)\;  L^{bc}_{m+n} +
\frac{c(\lambda)}{12}\; m\;(m^2 -1)\; \delta_{m+n,0}\cr
[L^{bc}_m; U_n] &=& -n\;U_{m+n} + \frac{Q}{2}\;m\,(m+1)\;\delta_{m+n,0}\cr
[U_m ; U_n] &=& \epsilon\;m\;\delta_{m+n,0}\;\;\;\;,\label{ghostviralg}
\eea
where the central charge $c(\lambda)$ and background charge $Q$ are
\be
c(\lambda) = -2 +12\,\lambda\,(1-\lambda) = 1 - 3\, Q^2\;\;\;\;,\;\;\;\; Q = 1
-2\,\lambda\;\;\;\;,
\ee
while the vacuum conformal dimension and ghost charge are determined to be
\bea
\Delta^{bc}_0 &=& \frac{1}{2}\, (p_0 -\lambda)\;(p_0 +\lambda-1) =
\frac{1}{2}\, u_0\,( u_0 + Q )\cr
u_0 &=& p_0 + \lambda -1\;\;\;\;.
\eea
We remark that the minimal conformal dimension state (``vacuum") is $|0>_{\epsilon_0}$;
only when $\epsilon_0=0$  the state $|+>\equiv c_0\,|0>_0 = |0>_1$ is degenerate with
$|->\equiv |0>_0$, both with dimension $\Delta^{bc}_0 = \frac{\lambda}{2}\; (1-\lambda)$.
However, the $SL(2,\br)$ invariant vacuum $|1> = |0>_{1-\lambda}$ does not coincide
(except when $\lambda=1$) with any of them.
In any case, as it happens with the fermions, the identifications
\be
|0>_{p_0 +m}\equiv \left\{ \begin{array}{lcr}b_{p_0+m}\dots b_{p_0 -1}\;|0>_{p_0}&,&
m\in\bz^-\cr c_{-p_0-m+1}\dots c_{-p_0}\;|0>_{p_0} &,&m\in\bz^+\end{array}\right.
\label{identifications}
\ee
allow us to take $p_0=\epsilon_0$ with no loss of generality.

\subsection{Commuting $\beta-\gamma$ systems}

Similarly to what we have just considered, we can take now a pair of spinorial but
commuting fields $(\beta,\gamma)$ where, this time, $\lambda$ is half-integer.
With the replacements $(b,c)$ by $(\beta,\gamma)$, $(b_p , c_{-p})$ by
$(\beta_p, \gamma_{-p})$, etc, things go similarly, so we just point out the main
differences.
In the first place the commutation relation
\be
[\gamma_{-p};\beta_q ] = \delta_{p,q}\;\;\;,\;\;\; p,q\in\bz_{\epsilon_0}
\ee
holds.
The cases of interest in superstring theory are $\epsilon_0=0$ and
$\epsilon_0 =\frac{1}{2}$, which follow the modding of the world-sheet SUGRA
transformation parameters in the R and NS sectors respectively.
A reference state $|0>_{\pi_0}$ is defined as in (\ref{ghostvacuum}) but, since relations
such as (\ref{identifications}) do not exist, each one of them defines different
representations or $u_0 =1 - \lambda-\pi_0$ ``pictures" \cite{yost}.
For $\lambda=\frac{3}{2}$, for example, we have $u_0=-\frac{1}{2}-\pi_0$; the vacuum states
are the states with $\pi_0= \frac{1}{2}$ and $\pi_0=0$ in the NS and R sectors respectively,
the commonly used ``$-1$" and ``$-\frac{1}{2}$" pictures.
It is worth noting that, due to the bosonic character of the
operators, the R vacuum state $|0>_0$ is infinitely degenerated (instead of doubly) with the
set of states $\gamma_0{}^{m_0}\,|0>_0$ with $m_0=0,1,2,\dots$.
On the other hand, the $SL(2,\br)$ invariant state is instead identified with the NS state
$|0>_{1-\lambda}$, the reference state of the ``0" picture.

The parameters and constants of the conformal-ghost algebra (\ref{ghostviralg}) (with
$\epsilon =-1$) are
\bea
c(\lambda) &=& 2  - 12\,\lambda\,(1-\lambda) = -1 + 3\,
Q^2\;\;\;\;,\;\;\;\; Q = 2\,\lambda-1\cr
\Delta^{\beta\gamma}_0 &=& \frac{1}{2}\, (\pi_0-\lambda)\;(1-\lambda-\pi_0)
= -\frac{1}{2}\, u_0\,( u_0 + Q )\cr u_0 &=& 1 - \lambda-\pi_0\;\;\;\;.
\eea
Of particular interest is the combined system of a $\lambda=2$ $b$-$c$ and
$\lambda=\frac{3}{2}$ $\beta$-$\gamma$, present in the superstring.
In this case we have
\be
c^g = -26 + 11 = -15 \;\;\;,\;\;\; \Delta^g_0  = -1 + \left\{
\begin{array}{r}\frac{3}{8} \cr\frac{1}{2}\end{array}\right. = \left\{
\begin{array}{rl}-\frac{5}{8}&,\;\;if\;\; R\cr -\frac{1}{2}&,\;\; if\;\;
NS\end{array}\right.\;\;\;\;.\label{ccsg}
\ee
\vfill\eject

\section{Partition functions}
\cleqn

Here we list the relevant partition functions of the various systems considered in
precedence. We introduce the variables $q\equiv e^{i2\pi\,\tau}=e^{-2\pi\,t}$ and
$z\equiv e^{i2\pi b}$, $V_n$ stands for the $\br^n$-volume, the prime in a trace symbol
means omitting the zero mode sector and the modular functions $\eta(\tau)$,
$\theta_{\alpha\beta}(\nu,\tau)$ are as in chapter $7$ of \cite{polcho1} (see i.e.
\cite{fay} for more),
\bea
\eta(\tau)&\equiv& q^\frac{1}{24}\; \prod_{m=1}^\infty\;(1-q^m)
\cr
Z^{2a}_{2b}(\tau) &\equiv& \frac{\vartheta\left[\matrix{a\cr b}\right]
(0;\tau)}{\eta(\tau)} = \frac{\vartheta_{2a,2b}(0;\tau)}{\eta(\tau)}\cr
&=&z^a\;q^{\frac{a^2}{2}-\frac{1}{24}}\; \prod_{m=1}^\infty\;(1 + z\,
q^{m-\frac{1}{2}+a})\, (1+ z^{-1}\, q^{m-\frac{1}{2}-a})\label{modfun}
\eea
They satisfy the modular properties
\bea
\eta(\tau)&=& (-i\,\tau)^{-\frac{1}{2}}\;\eta(-\tau^{-1}) = e^{-i\frac{\pi}{12}}\;
\eta(\tau+1)\cr
\vartheta\left[\matrix{a\cr b}\right](\nu;\tau) &=&
(-i\,\tau)^{-\frac{1}{2}}\; e^{i\pi(2ab - \tau^{-1}\nu^2)}\;
\vartheta\left[\matrix{b\cr -a}\right](-\tau^{-1}\nu;-\tau^{-1})\cr
&=&e^{i\pi a(a+1)}\;\vartheta\left[\matrix{a\cr b-a-\frac{1}{2}}\right](\nu;\tau +1)\cr
Z^{2a}_{2b}(\tau) &=& e^{i2\pi ab}\;Z^{2b}_{-2a}(-\tau^{-1}) =
e^{i\pi (a^2+ a + \frac{1}{12})}\;Z^{2a}_{2b-2a-1}(\tau+1)
\label{modtransf}
\eea
We get
\bea
Z^{X}(\tau) &\equiv& tr\; q^{L^X_0 -\frac{1}{24}} =\left\{\begin{array}{ll} V_1\;
(2\pi\,l)^{-1}\;t^{-\frac{1}{2}}\;\eta(\tau)^{-1}\;&\;,\;\;\; NN\cr
e^{-T_s\,\Delta x^2\,t}\;\eta(\tau)^{-1}\;&\;,\;\;\; DD\cr
(Z^0_1(\tau))^{-\frac{1}{2}}\;&\;,\;\;\;ND/DN\end{array}\right.\cr
& &\cr Z^{\psi}(\tau) &\equiv& tr\; e^{i\pi\beta F}\; q^{L^\psi_0
-\frac{1}{48}} = \left( Z^\alpha_\beta(\tau)\right)^{\frac{1}{2}}
\;\;\;\;,\;\;\;\; \alpha = \left\{\begin{array}{l}0\;\;\;,\;\;\;NS\cr 1\;\;\;,\;\;\;
R\end{array}\right.\;\; ,\;\;\beta = \left\{\begin{array}{l}0\;\;\;,\;\;\; AP\cr
1\;\;\;,\;\;\;P\end{array}\right.\cr
& &\cr
Z^{bc}(\tau)&\equiv&\left\{\begin{array}{ll}
tr_{p_0}\; q^{L^{bc}_0 -\frac{c(\lambda)}{24}}\; z^{U^{bc}_0 -\lambda +\frac{1}{2}} =
Z^{2p_0 +1}_{2b}(\tau)&\cr tr'_{p_0}\; q^{L^{bc}_0
-\frac{c(\lambda)}{24}}\;(-)^{U^{bc}_0 -\lambda +1} \equiv
\frac{Z^1_{2b}(\tau)}{2\,\cos \pi b}|_{b=\frac{1}{2}} =
\eta(\tau)^2&,\;\;\; p_0\in\bz\end{array}\right.\cr
Z^{\beta\gamma}(\tau)&\equiv&\left\{\begin{array}{ll} e^{i\pi(\pi_0 -\frac{1}{2} )}\;
tr_{\pi_0}\; q^{L^{\beta\gamma}_0
-\frac{c(\lambda)}{24}}\; z^{U^{\beta\gamma}_0 +\lambda -\frac{1}{2}} =
(Z^{1- 2\pi_0 }_{1-2b}(\tau))^{-1}&,\;\;\; (\pi_0,b)\notin
\bz\times\bz \cr
tr'_{\pi_0}\;q^{L^{\beta\gamma}_0 -\frac{c(\lambda)}{24}}\; \equiv
2\,\sin \pi b\; (Z^1_{1-2b}(\tau))^{-1}|_{b=0} = \eta (\tau)^{-2}&,\;\;\;\;\;\;\;\;\;
\pi_0\in\bz\cr
\end{array}\right.\cr
Z^Z(\tau) &\equiv& tr_{\bar\nu}\; q^{L^Z_0 -\frac{1}{12}} = \left\{\begin{array}{ll}
\frac{V_2}{2\,\pi\,\theta}\, e^{i\pi(\frac{1}{2}-\bar\nu)}\; (Z^{1-2\bar\nu}_1(\tau))^{-1}
&\;,\;\;\;\bar\nu\neq 0 \cr \frac{V_2}{8\pi^2\alpha'\cos^2\varphi_0{}\,t} \;\eta(\tau)^{-2}
&\;,\;\;\; \bar\nu =0\end{array}\right.\cr
Z^\Psi(\tau) &\equiv& tr_{r_0}\; q^{L^\Psi_0 -\frac{1}{24}}\; z^{-J^\Psi_0} =
Z^{1-2r_0}_{2b}(\tau)\label{pf}
\eea
\vfill\eject

\end{document}